\documentclass[pdflatex,sn-mathphys-ay]{sn-jnl}

\usepackage{graphicx}%
\usepackage{multirow}%
\usepackage{amsmath,amssymb,amsfonts}%
\usepackage{amsthm}%
\usepackage{mathrsfs}%
\usepackage[title]{appendix}%
\usepackage{xcolor}%
\usepackage{textcomp}%
\usepackage{manyfoot}%
\usepackage{booktabs}%
\usepackage{algorithm}%
\usepackage{algorithmicx}%
\usepackage{algpseudocode}%
\usepackage{listings}%

\usepackage{orcidlink}
\usepackage{multirow}
\usepackage{makecell}
\usepackage{xspace}
\usepackage{caption}
\usepackage{subcaption}

\theoremstyle{thmstyleone}%
\theoremstyle{thmstyletwo}%

\theoremstyle{thmstylethree}%

\usepackage{xspace}

\newcommand{\ssumo}{\textsf{SSSUMO}\xspace}

\begin{document}

\title{\ssumo: Real-Time Semi-Supervised Submovement Decomposition}


\author*[1]{\fnm{Evgenii} \sur{Rudakov}\orcidlink{0009-0003-8204-302X}}\email{evgenii.rudakov@helsinki.fi}

\author[2,3,4]{\fnm{Jonathan} \sur{Shock}\orcidlink{0000-0003-3757-0376}}\email{jonathan.shock@uct.ac.za}

\author[5]{\fnm{Otto} \sur{Lappi}\orcidlink{0000-0002-7996-7665}}\email{otto.lappi@helsinki.fi}

\author[5,1]{\fnm{Benjamin Ultan} \sur{Cowley}\orcidlink{0000-0001-8828-2994}}\email{ben.cowley@helsinki.fi}

\affil[1]{\orgdiv{Faculty of Educational Sciences}, \orgname{University of Helsinki},
           \orgaddress{\city{Helsinki}, \country{Finland}}}

\affil[2]{\orgdiv{Department of Mathematics and Applied Mathematics}, \orgname{University of Cape Town},
          \orgaddress{\city{Cape Town}, \country{South Africa}}}

\affil[3]{\orgname{INRS}, \orgaddress{\city{Montreal}, \country{Canada}}}

\affil[4]{\orgname{NiTheCS}, \orgaddress{\city{Stellenbosch}, \country{South Africa}}}

\affil[5]{\orgdiv{Cognitive Science, Department of Digital Humanities}, \orgname{University of Helsinki},
          \orgaddress{\city{Helsinki}, \country{Finland}}}


\abstract{
This paper introduces a \ssumo, semi-supervised deep learning approach for submovement decomposition that achieves state-of-the-art accuracy and speed. While submovement analysis offers valuable insights into motor control, existing methods struggle with reconstruction accuracy, computational cost, and validation, due to the difficulty of obtaining hand-labeled data. We address these challenges using a semi-supervised learning framework. This framework learns from synthetic data, initially generated from minimum-jerk principles and then iteratively refined through adaptation to unlabeled human movement data. Our fully convolutional architecture with differentiable reconstruction significantly surpasses existing methods on both synthetic and diverse human motion datasets, demonstrating robustness even in high-noise conditions. Crucially, the model operates in real-time (less than a millisecond per input second), a substantial improvement over optimization-based techniques. This enhanced performance facilitates new applications in human-computer interaction, rehabilitation medicine, and motor control studies. We demonstrate the model's effectiveness across diverse human-performed tasks such as steering, rotation, pointing, object moving, handwriting, and mouse-controlled gaming, showing notable improvements particularly on challenging datasets where traditional methods largely fail. Training and benchmarking source code, along with pre-trained model weights, are made publicly available at \href{https://github.com/dolphin-in-a-coma/sssumo}{\texttt{github.com/dolphin-in-a-coma/sssumo}}.}




\maketitle

\section{Introduction}
\label{sec:intro}

Human motor actions are not fully continuous, and a complex motion (e.g. reaching) can be represented as a summation of simpler constituents known as submovements \citep{hogan1984, neural_evidence1, neural_evidence2, review_paper1, peak_detector}. In each submovement, the end-effector (e.g. arm) is then typically found to exhibit a minimum-jerk principle minimizing the rate of change of acceleration, and to follow a smooth, near-straight path with a characteristic unimodal, bell-shaped velocity profile that scales with movement amplitude \citep{hogan1984}. However, velocity profiles for movements performed under constraint, e.g. speeded tasks, often deviate from the ideal, likely reflecting discrete corrective submovements (first noted by \citet{woodworth1899}). We can think of each submovement as a discrete, error-initiated correction to the motion trajectory \citep{active_inference, markkula_control}. This framework allows us to infer sensorimotor system strategies, because if the nervous system corrects course at discrete times, detecting the resulting submovements elucidates the relationship between task conditions and motor action planning.

In this work, we introduce a novel Semi-Supervised approach to SUb-MOvement decomposition (\ssumo) that leverages both synthetic data and unlabeled, human-generated data. We benchmark our \ssumo model against the most widely-used approaches on a set of different hand-performed motor activities, demonstrating a novel solution to the challenges of submovement detection and consequent limitations of existing methods.


Detecting submovements presents several challenges. First, human movement data are often \textit{noisy} and \textit{highly variable}. Second, \textit{overlapping submovements} can appear as asymmetric or multi-peak movements, typical in tasks with high precision demands. Decomposing this overlap is an ill-posed problem: multiple configurations of smaller components could plausibly fit any given movement profile \citep{spurious1}. Third, we lack \textit{labeled datasets} of submovements. In contrast to applications like image recognition, a human annotator cannot directly observe labels for each submovement without an algorithmic decomposition \citep{spurious1}. Ground-truth labels for submovements are thus sparse and estimate-driven, which complicates supervised learning approaches.

A range of techniques have been developed to parse these underlying submovement structures, from heuristic methods \citep{peak_detector, wavelet} to computationally intensive optimization approaches \citep{spurious1, spurious2, updated_scattershot, updated_scattershot2}. Heuristics may be, for instance, that each velocity peak corresponds to a distinct submovement -- these methods, though computationally efficient, often fail when submovements interact, such as two close submovements with a single velocity peak. While optimization algorithms exist that are guaranteed to find the global minimum \citep{spurious1}, their running time grows exponentially with the number of submovements making them impractical for large datasets. Furthermore, finding global minima may lead to overfitting with noisy or low-resolution velocity signals, or when starting with incorrect assumptions. Polynomial-time optimization algorithms, while faster, are probabilistic in nature and prone to getting stuck at local minima \citep{spurious2, updated_scattershot}, sometimes yielding reconstruction quality inferior to heuristic methods. Moreover, until now, previously described solutions haven't been released as open-source software or thoroughly benchmarked, slowing down the adoption and progress in the field.

The key idea behind our \ssumo algorithm is to harness prior knowledge of submovement characteristics to generate synthetic motion profiles with known (labeled) submovements, and use these to bootstrap learning on real human movement data where submovement labels are unknown. Our approach consists of two main stages:
\begin{enumerate}
    \item \textbf{Initial Stage:} 
    \begin{itemize}
        \item \textbf{Synthetic Pre-training:} 
        Pre-train a decomposition model on initial synthetic velocity profiles with known submovements. 
        \end{itemize}
    \item \textbf{Iterative Refinement Stage:} The following steps are repeated until convergence:
    \begin{itemize}
        \item \textbf{Pseudo-Labeling:} Infer pseudo-labels on real data and estimate submovement parameter distributions.
        \item \textbf{Synthetic Refinement:} Generate refined synthetic data based on estimated distributions.
        \item \textbf{Model Re-training:} Re-train the model on the updated synthetic data.
    \end{itemize}
\end{enumerate}

By alternating learning between synthetic and unlabeled real data, this semi-supervised approach benefits from the best of both worlds: the abundance and clarity of synthetic data and the ecological validity of real data. A major advantage of \ssumo is that initial training in the fully-labeled synthetic domain provides a strong `first guess', that is then gradually adapted to handle real-world complexity without requiring any hand-labeled ground-truth in human data. This strategy circumvents the need for large labeled training sets – instead, the large unlabeled dataset itself guides the model. Moreover, by regenerating synthetic examples that match the real-data characteristics, the method continuously improves its realism and accuracy, in a way akin to expectation-maximization in model space \citep{self_training1, self_training2, self_training3}.

Our results show that \ssumo's semi-supervised training loop outperforms other solutions while operating in real-time, generalizes to all tested activities, is robust to various noise conditions in the velocity signal, and may be readily adapted to new activities. We open-source both the code and weights to facilitate future advancements in the field.

\section{Related Work}

\subsection{Evidence for Submovements in Human Motor Control}
A growing body of research suggests that complex hand movements are composed of smaller submovements or motion primitives \citep{hogan1984, neural_evidence1, neural_evidence2, review_paper1, peak_detector}. These submovements manifest as brief velocity pulses or adjustments at frequencies of 2-3 Hz \citep{review_paper1}, reaching as high as 8-12 Hz for fast-paced tasks such as handwriting \citep{fast_writting, fast_neuro, high_freq1, high_freq2}. Empirical evidence comes from kinematic analyses showing that even simple reaching motions often display multiple peaks in their speed profiles rather than a single smooth bell-shaped velocity curve \citep{Milner1992}. These multi-peaked velocity traces are commonly interpreted as indicating sequential submovements, especially in tasks requiring precision.

Neurophysiological findings further bolster this view. Studies have identified neural activity patterns time-locked to submovement events, revealing that the motor cortex exhibits discontinuous activation patterns that correlate with kinematic fluctuations in the movement trajectory \citep{neural_evidence1, neural_evidence2}. For example, brain recordings in humans have revealed event-related potentials in the supplementary motor area (SMA) that coincide with each submovement during continuous tracking tasks \citep{neural_SMA}. Such neural correlates lend credence to motor control models that hypothesize that movements are generated as a series of discrete units.

Classical theories of goal-directed motion, like iterative correction models, explicitly propose that an initial ballistic move toward a target is followed by one or more corrective submovements to refine accuracy \citep{Meyer1988}. The minimum intervention principle suggests that the nervous system only makes corrections when deviations from the intended trajectory exceed certain thresholds, naturally leading to a discrete, intermittent control strategy \citep{Todorov2002}.

Nevertheless, the submovement hypothesis is not without its critics. Some researchers argue that the submovement framework is overly simplistic and that movements are better understood as continuous processes \citep{opposite_view2, opposite_view3} 
These critics suggest that peaks in velocity profiles originate from inherent characteristics of the motor system rather than from discrete adjustments.

In sum, despite existing criticisms, converging evidence from behavioral kinematics, cortical activity, and computational modeling indicates that submovements are a real and integral aspect of human motor control, serving as plausible building blocks for constructing longer, complex actions. For the purposes of our research, we accept submovements as a valid concept based on substantial evidence.


\subsection{Applications of Submovement Decomposition}
The ability to decompose movements into their constituent submovements has proven valuable across multiple domains. One prominent application is stroke rehabilitation, where submovement analysis provides quantitative metrics of motor impairment and recovery. Stroke survivors often execute reaching movements in a segmented, jerky fashion - essentially stringing together many small submovements due to weakened or uncoordinated control. As patients recover motor function through therapy, their movements tend to grow smoother and more efficient, which corresponds to submovements that are fewer in number, longer in duration, and higher in peak speed. In essence, recovery involves merging or `blending' submovements that were formerly separate, resulting in more continuous motion \citep{hogan_stroke}.
Tracking these changes has clinical value: a decrease in the number of submovements (or an increase in their overlap) serves as an objective sign of improved motor performance and coordination. Researchers have leveraged submovement decomposition to evaluate patient progress, adjust rehabilitation strategies, and even as a basis for therapeutic interventions that encourage smoother motions \citep{hogan_stroke, stroke_recovery2}.

Beyond rehabilitation, submovement decomposition has been applied to skill analysis in domains such as handwriting and motor learning. In handwriting analysis, individual pen strokes or segments of cursive writing can be interpreted as overlapping submovements that superimpose to form complex letter trajectories \citep{Plamondon1995}. Computational models of handwriting have successfully treated the pen’s velocity profile as a sum of lognormal or bell-shaped submovement pulses, reproducing the characteristic shapes of written characters \citep{Plamondon2006}. Decomposing handwriting in this way has helped in understanding the neuromuscular synergies involved in fine motor tasks and in designing algorithms for character recognition.


Similarly, researchers have studied how submovements manifest in motor learning and skill acquisition \citep{motor_learning, motor_learning2, motor_learning3, motor_learning4}. Novice learners tend to make movements that are halting and require frequent mid-course corrections – essentially more submovements – whereas expert performance is smoother with corrections blended in or eliminated \citep{motor_learning3}. For example, in a point-to-point reaching task, beginners overshoot or undershoot and then perform additional submovements to reach the target, but with practice, these secondary submovements reduce or disappear. Quantitative analyses confirm that as people gain skill, their movements involve longer and larger submovements executed with greater overlap, which in turn yields smoother motion trajectories \citep{leg_obstacle}. 
This pattern holds for learning in more complex tasks, such as mouse-controlled game-playing \citep{game1, game2}. 
Interestingly, analogous patterns also emerge during early ontogeny \citep{kids}.


Submovement decomposition has also contributed to the development of adaptive human-computer interfaces \citep{autogain, interface1, interface2} and motor control models \citep{markkula_control, control1, control2, control3, control4}. For instance, segmentation of computer mouse movements into submovements has enabled the development of an adaptive gain control algorithm that automatically adjusts mouse cursor gain with shorter adaptation time while performing comparably to high-end industry solutions \citep{autogain}. Similarly, the decomposition of steering motion signals into submovements has led to the development of a driver control model that enhances the accuracy of steering signal prediction \citep{markkula_control}.



\subsection{Optimization-Based Decomposition Methods}

Most traditional approaches to submovement decomposition pose it as a parameter estimation or optimization problem. The decomposition is formulated as an inverse problem. One assumes that an observed end-effector trajectory is generated by the superposition of several underlying primitive motions, each following a stereotyped shape \citep{spurious1}. The task is then to find a set of such primitives (with specific displacements, durations, start times, etc.) that sum up to reconstruct the original movement as closely as possible.

The Branch and Bound algorithm is one of the methods that are proven to find the global minima \citep{spurious1}. However, it requires exponential time and takes several hours to decompose a signal containing just three submovements \citep{spurious2}.

The most influential method is the polynomial-time Scattershot optimization algorithm \citep{spurious2}. This probabilistic optimization method iteratively increases the number of submovements fitted to the velocity signal until the fit error—quantified by the mean absolute percentage error (MAPE)—drops below a predetermined threshold. For each candidate number of submovements, the algorithm randomly samples starting positions, displacements, and durations from predefined ranges. The original work argued that ten independent runs were sufficient to approach a globally optimal solution.


This approach has the advantage of making minimal assumptions about the signal aside from the general shape, duration, and displacement of submovements, and it performs well when the number of submovements is relatively low (as is typical in shorter recordings). However, its computational complexity increases rapidly with the number of submovements. Specifically, the time complexity of a single fit grows as a third-order polynomial with respect to the number of submovements, and the likelihood of converging on the globally optimal solution decreases as more submovements are introduced. Aside from runtime issues, optimization-based approaches exhibit notable sensitivity to measurement noise and their own modeling assumptions. Empirically, when motion data is even slightly noisy, an optimizer may misinterpret minor fluctuations as additional submovements that are not actual movement corrections.

Scattershot is still widely adopted \citep{robot, vr} and has influenced updated methods, which are also polynomial-time, probabilistic, and susceptible to the same issues, and these have not gained significant traction in the field \citep{updated_scattershot, updated_scattershot2}.

\subsection{Heuristic and Linear-Time Approaches}
Given the computational cost of optimization methods, researchers have also explored faster, heuristic techniques for submovement decomposition. These approaches typically operate in linear time with respect to the number of submovements, making them attractive for real-time or large-scale processing. A common strategy is to identify submovements by detecting simple patterns in the kinematics, such as scanning the velocity profile for local minima and maxima \citep{peak_detector}. For instance, one might smooth the speed curve of the hand movement and then mark each sequence of `minimum–maximum–minimum' as a candidate submovement unit. Other heuristics utilize higher-order derivatives; for example, counting zero-crossings of jerk or snap can indicate an underlying intermittent control strategy \citep{motor_learning}.
More sophisticated methods include wavelet-based approaches \citep{wavelet} to detect submovement onsets as singularities in the wavelet transform.

However, the speed gained by heuristic decomposition comes at the cost of accuracy and flexibility. Such methods make strong assumptions about how submovements manifest in the observed signals. For example, a peak-detection method assumes that submovements do not heavily overlap in time; if two submovements overlap without a discernible dip in speed, they may be erroneously merged. Conversely, noise or minor fluctuations can lead to over-segmentation, where spurious peaks are mistaken for genuine submovements. The reliance on fixed thresholds and simplistic decision rules means that performance can degrade rapidly when the movement deviates from idealized conditions.

\subsection{Incorporation of Machine Learning and Synthetic Data}
In recent years, data-driven techniques have been applied to submovement decomposition to address some limitations of purely model-based methods. One direction has been supervised learning: using labeled examples of decomposed movements to train a model that can decompose new movements. For instance, \citet{liao_characterizing_nodate} decomposed a set of human 3D reaching trajectories using a minimum-jerk optimization method, then trained an artificial neural network (ANN) to predict submovement parameters from hand-crafted features of the motion. This approach demonstrated that once provided with decomposition `ground truth' (obtained offline), a learned model could quickly predict submovements for new trajectories in a feed-forward manner, significantly speeding up the decomposition process. However, a drawback was the need to hand-craft input features and the reliance on an initial decomposition algorithm to provide training labels, whose performance served as a limiting factor.

Unsupervised and statistical learning methods have also been explored to infer submovements or movement primitives without explicit labels. Some studies use cluster analysis on segments of movement to discover common shapes. \citet{miranda} introduced the concept of "movement elements," which are sub-segments of motion identified in a data-driven way, and applied unsupervised clustering to these elements. 

Despite the limitations of both sophisticated optimization and simple heuristic approaches, the field of submovement analysis has yet to fully leverage modern data-driven techniques. A notable gap is the use of synthetic data to train machine learning models for submovement decomposition. Traditionally, synthetic movements—generated from a predefined set of submovement primitives—have been used primarily for benchmarking, as they provide a known ground truth for evaluating algorithms \citep{spurious2, wavelet}. However, while synthetic data serves as a convenient benchmark, it has rarely been used to \textit{develop} decomposition algorithms through learning.

One practical reason for the lack of progress in the development of learning-based methods is the lack of large labeled datasets of human movement broken down into submovements – after all, we cannot directly observe submovements in a person’s hand motion without already assuming a decomposition method. This creates a chicken-and-egg problem for supervised learning: there is no ground truth label for each timepoint of a human reach indicating which submovement it belongs to. 

Synthetic data offers considerable potential to overcome the current limitations of submovement decomposition. By simulating hand movements from known submovement primitives, one can produce virtually unlimited training examples with perfect knowledge of the underlying submovement parameters. These simulations can incorporate variations in speed, direction, submovement overlap, noise levels, and biomechanical constraints to emulate different real-world conditions.

Training a learning-based model on such a diverse synthetic dataset could imbue it with a far greater ability to generalize than any algorithm tuned on a small, homogeneous set of real movements. In essence, synthetic data can help break the dependency on rigid assumptions by allowing a model to learn the decomposition function from examples, rather than having it hard-coded.


\section{Human Motion Data}
\subsection{Datasets}
We searched specialized repositories (Zenodo, GitHub) and relevant scientific articles for open-access datasets on goal-directed activities. From these sources, we selected datasets based on the following criteria:

\begin{enumerate}
    \item \textbf{Task type} \newline
    The primary movement could be reduced to a goal-directed motion of the hand (or a hand-held object) from an initial position to one or more target positions. This permitted focusing on the gross trajectory rather than detailed finger arrangements or object orientation. This criterion provided tasks in three-dimensional space, as well as tasks constrained to a plane or along a single axis.

    \item \textbf{Direct position recording} \newline
    The dataset included direct spatial recordings (e.g., $x, y, z$ coordinates) of the hand or object, rather than inferring positions solely from inertial measurements. The sampling rate was at least 30~frames per second, ensuring adequate temporal resolution to reconstruct human-scale movement trajectories \citep{review_paper1}.

    \item \textbf{Number of subjects} \newline
    Each dataset included at least four subjects, thereby mitigating subject-specific idiosyncrasies. This ensured at least two subjects could be used for training and two for validation.

    \item \textbf{Licensing} \newline
    The dataset license permitted usage for research purposes.

\end{enumerate}

In total, we selected seven publicly available datasets, covering tasks with varying degrees of freedom ($D$): two 1D datasets, two 2D datasets, and three 3D datasets. This selection ensured a high diversity of goal-directed hand movements. 

Six public datasets were distributed under Creative Commons Attribution 4.0 International License (CC BY 4.0) licenses. The Handwriting dataset was made available under a custom research-only license.

\begin{figure}[!htbp]
    \centering
    
    \textbf{1D Tasks} \hspace{7cm} \textbf{2D Tasks}  \par\medskip
    \begin{subfigure}{0.22\textwidth}
        \includegraphics[width=\linewidth]{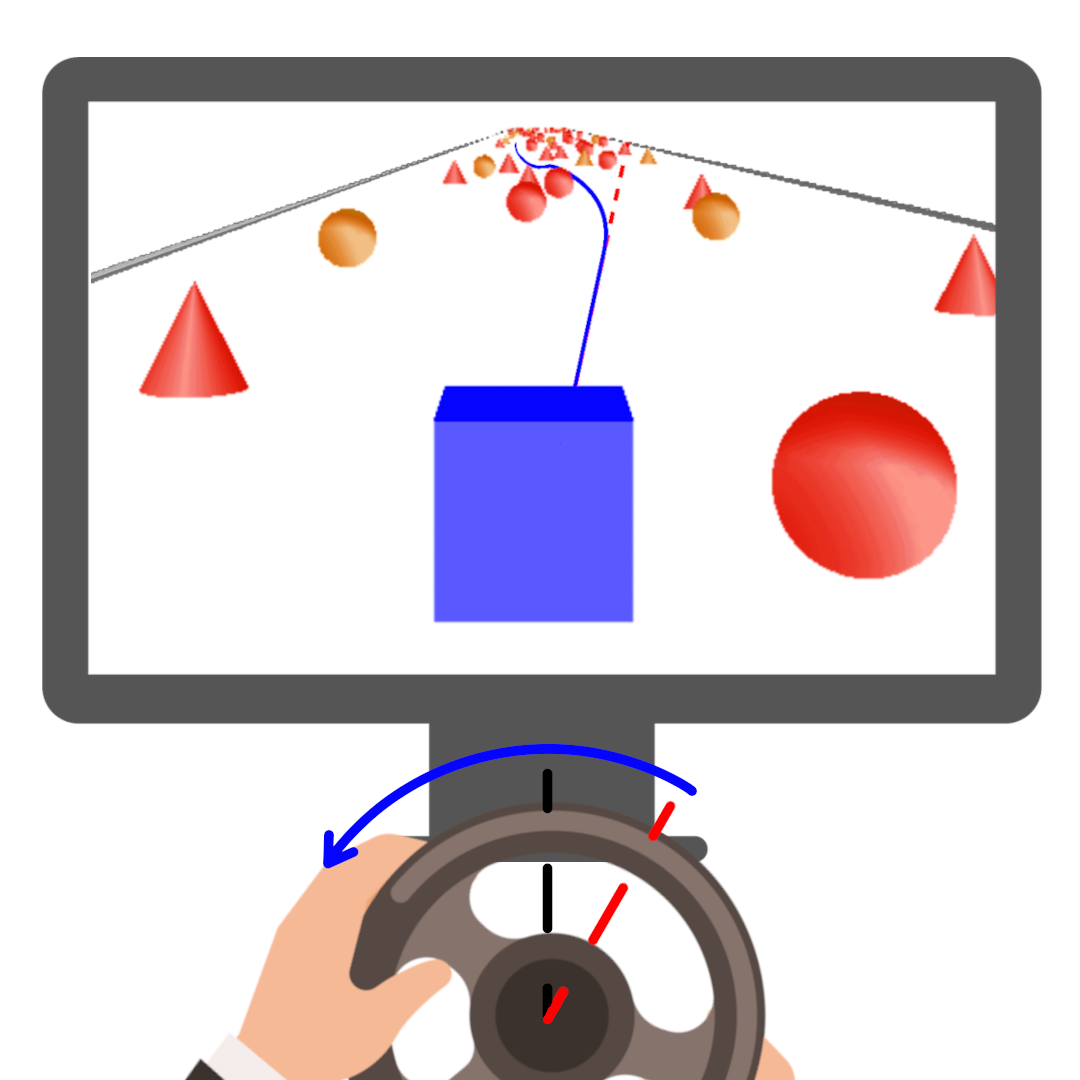}
        \caption{Steering 
        }
    
    \end{subfigure}
    \hfill
    \begin{subfigure}{0.22\textwidth}
        \includegraphics[width=\linewidth]{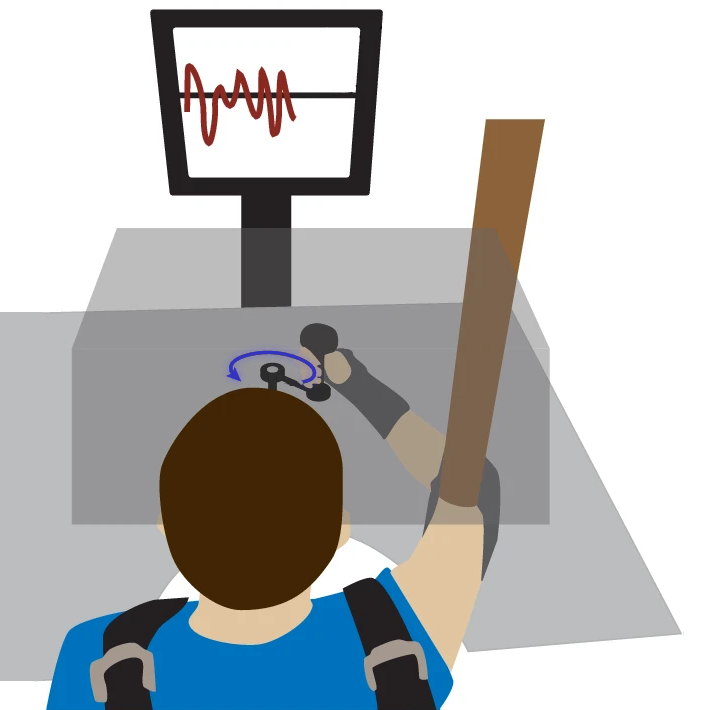}
        \caption{Crank Rotation 
        }
    \end{subfigure}
        \hfill
    \begin{subfigure}{0.22\textwidth}
        \includegraphics[width=\linewidth]{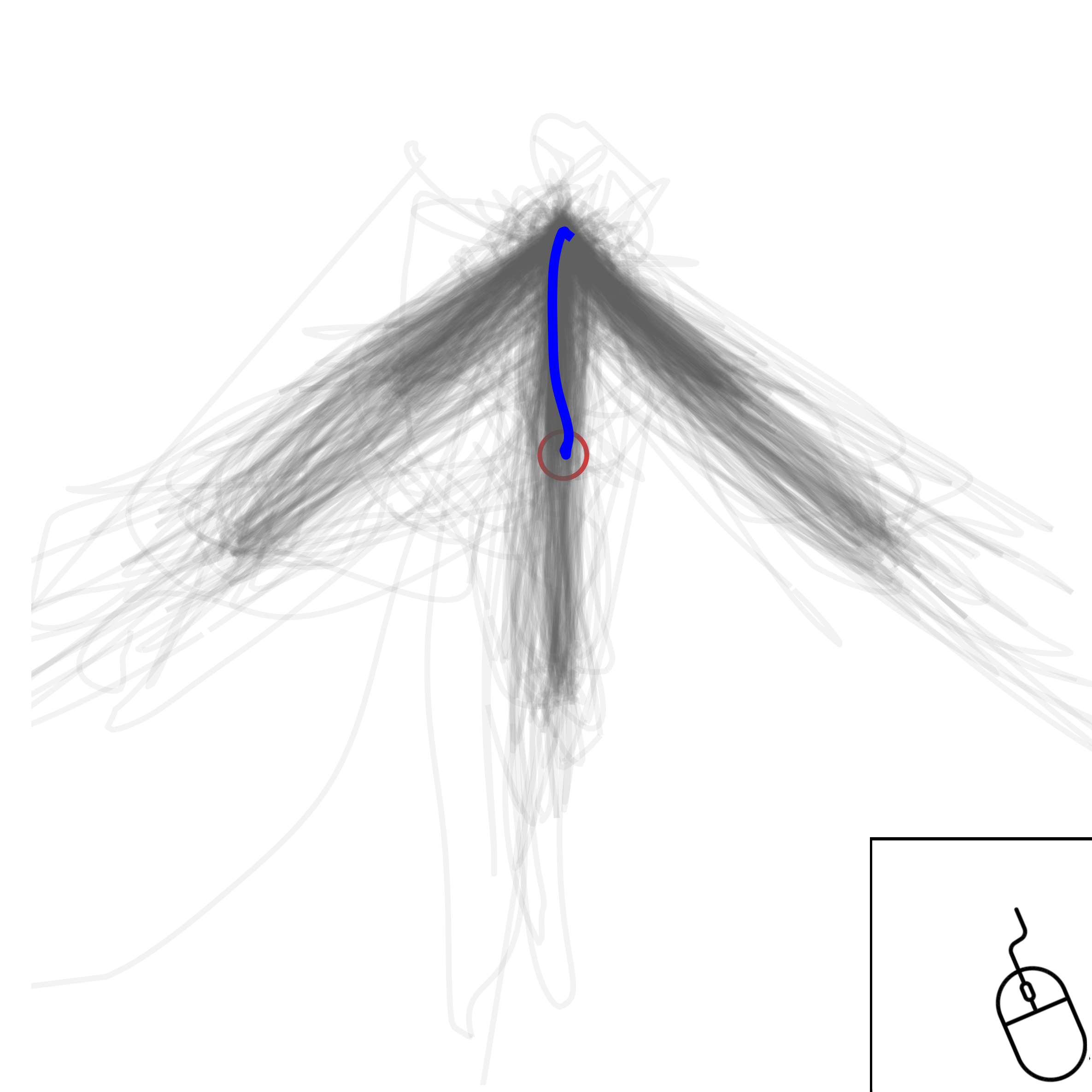}
        \caption{Fitts's Task 
        }
    \end{subfigure}
    \hfill
    \begin{subfigure}{0.22\textwidth}
        \includegraphics[width=\linewidth]{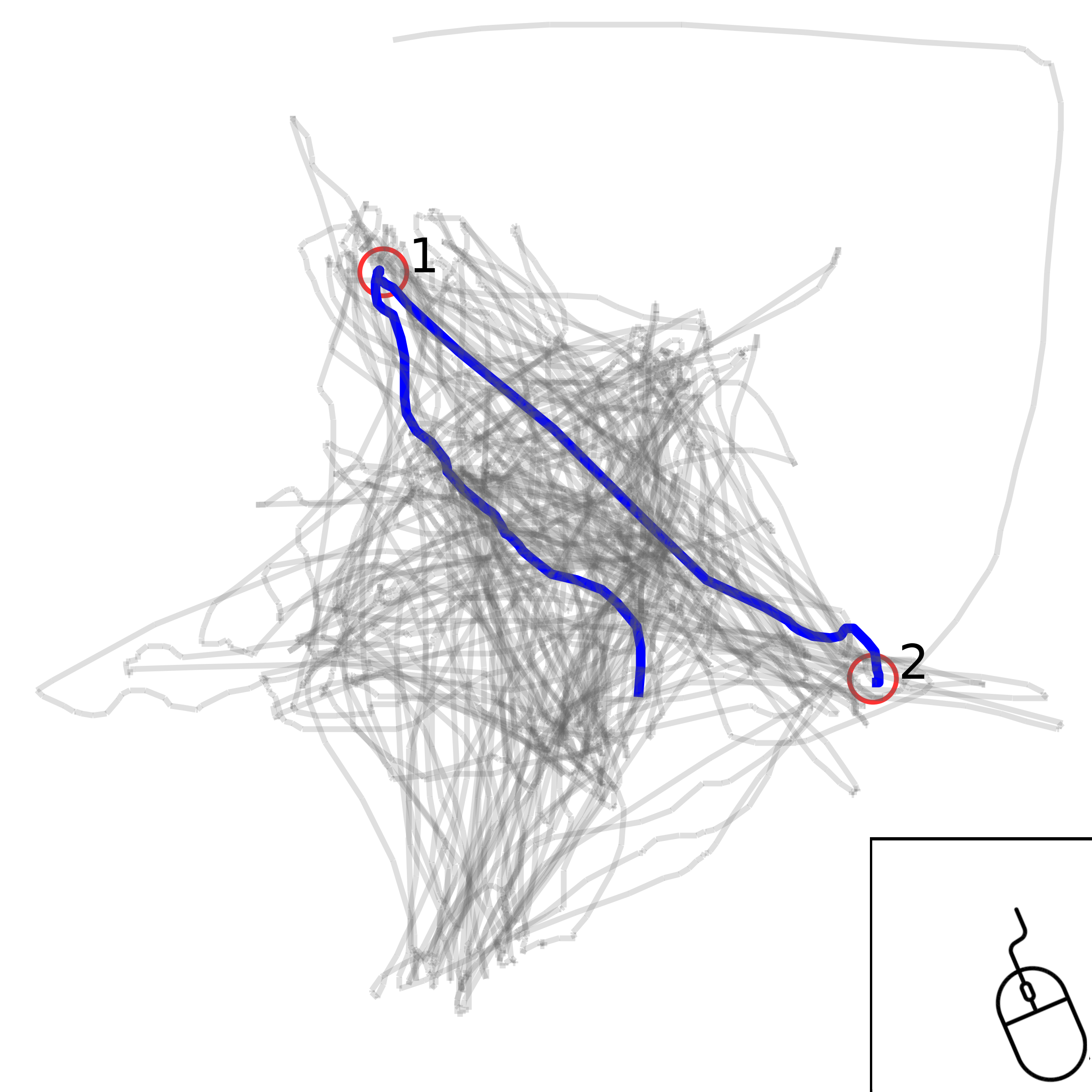}
        \caption{Whack-A-Mole 
        }
    \end{subfigure}

    \vspace{1em} 
    
    \textbf{3D Tasks} \par\medskip
    \begin{subfigure}{0.3\textwidth}
        \includegraphics[width=\linewidth]{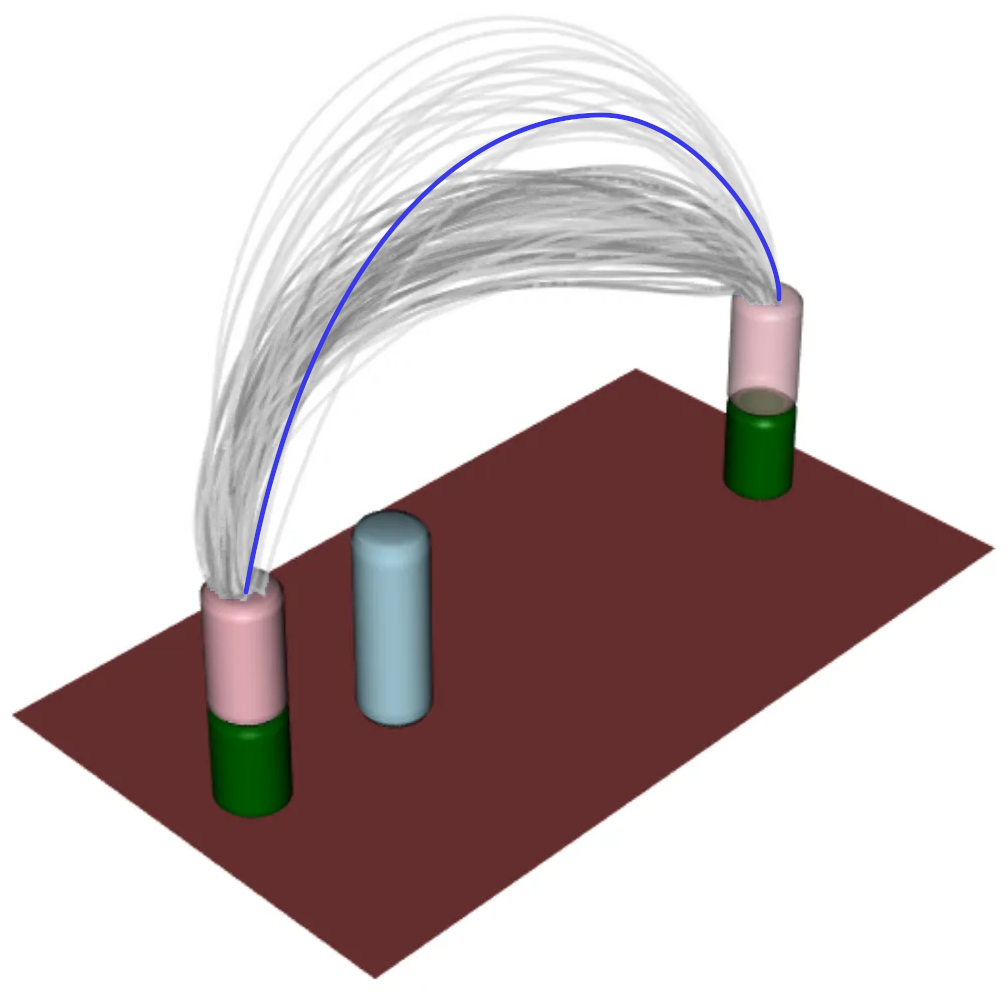}
        \caption{Object Moving 
        }
    \end{subfigure}
    \hfill
    \begin{subfigure}{0.3\textwidth}
        \includegraphics[width=\linewidth]{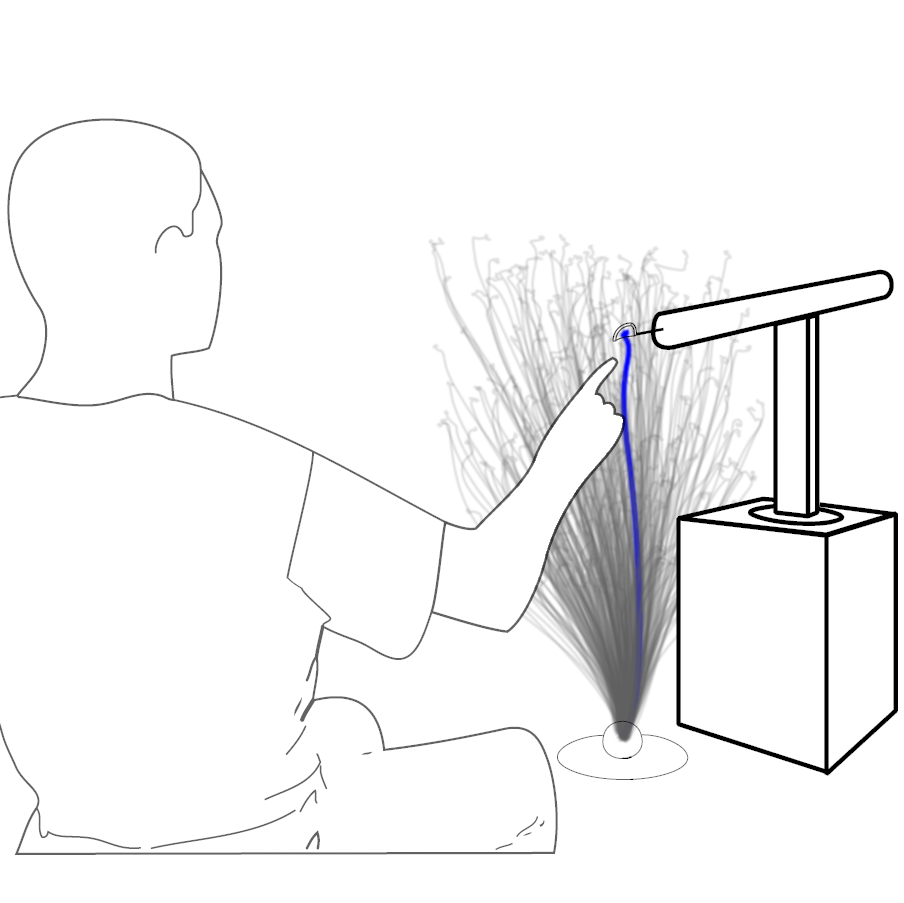}
        \caption{Pointing 
        }
    \end{subfigure}
    \hfill
    \begin{subfigure}{0.3\textwidth}
        \includegraphics[width=\linewidth]{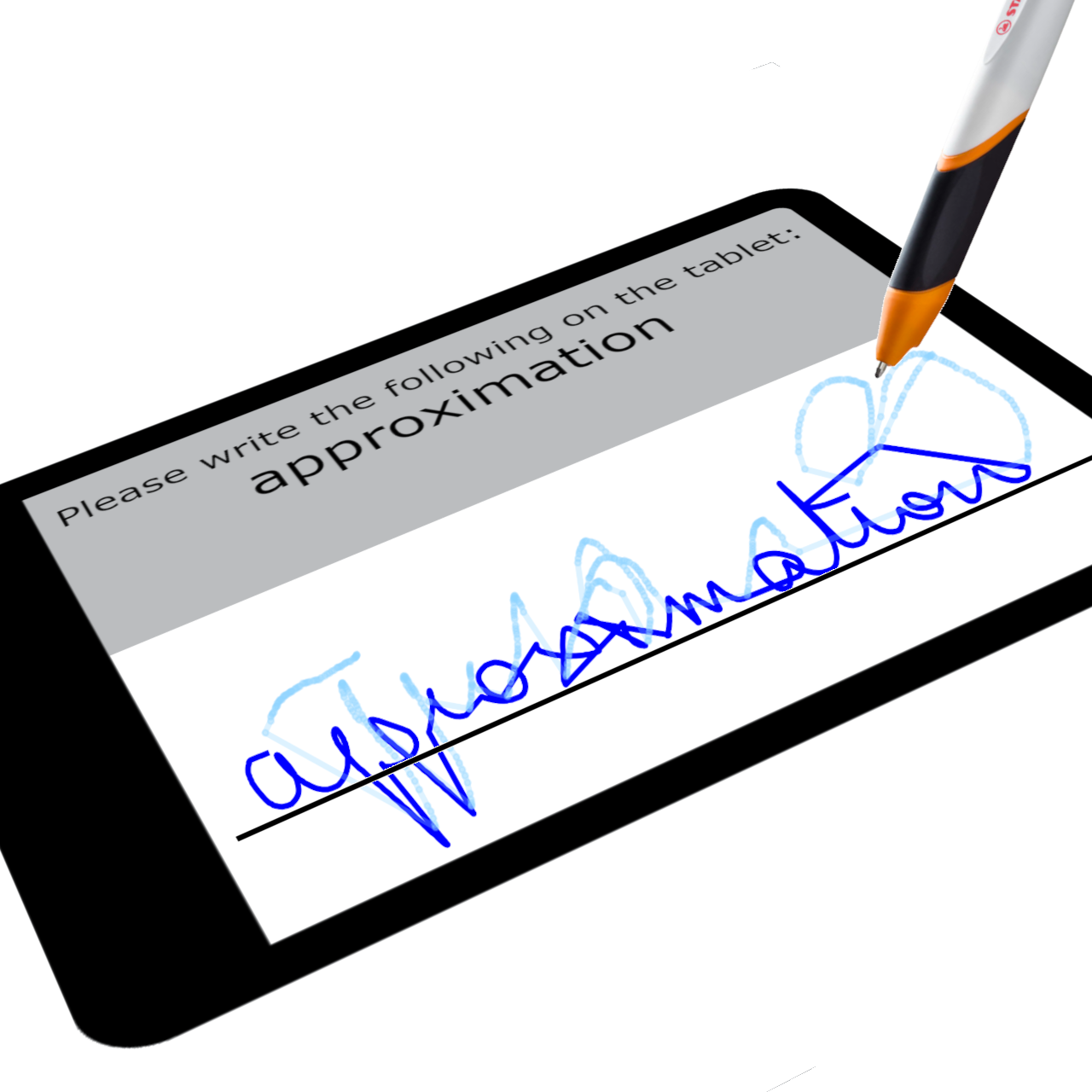}
        \caption{Handwriting 
        }
    \end{subfigure}
    
\caption{Visualizations of dataset tasks, grouped by dimensionality. The top row (a--d) features 1D and 2D tasks, and the bottom row (e--g) features 3D tasks. Where visible, semi-transparent gray lines show the paths of a randomly selected participant across all trials, while blue lines indicate the path taken in a single trial (or a portion thereof for longer tasks).
\textbf{(a)} A participant steers a blue cube using a steering wheel to avoid obstacles. The dashed red line shows the wheel's neutral center, indicating the predicted on-screen path if the wheel remains at its current position. A blue arrow denotes a movement segment, and the solid blue trace on the screen shows the actual path after wheel adjustments.
\textbf{(b)} A participant rotates a crank at a prescribed velocity, receiving real-time on-screen feedback about deviations from the target. The light brown canvas provides arm support. Adapted from \citet{tessari_brownian_2024}.
\textbf{(c)} Using a computer mouse, a participant rapidly moves from a fixed start area to one of six target areas. The red circle indicates the target area for the selected trial.
\textbf{(d)} In a `Whack-A-Mole' variant, a participant controls a computer mouse to quickly hit a pop-up `mole'. The red circles mark two consecutive positions of `moles' hit by the participant.
\textbf{(e)} A participant moves a pink cylinder toward a translucent pink target area while avoiding a blue cylinder. Adapted from \citet{raket_separating_nodate}.
\textbf{(f)} A participant uses their finger to point at a hollow quarter-sphere that appears in a new location at the start of each trial. Adapted from \citet{liao_characterizing_nodate}
\textbf{(g)} A participant writes a prompted word or sentence on a tablet using a smart pen. The dark blue trace shows the pen's on-screen strokes, and the light blue trace shows the midair path between strokes.}

    \label{fig:task_grouping}
\end{figure}

Table~\ref{tab:datasets} summarizes the datasets along with their key properties. Figure~\ref{fig:task_grouping} illustrates the mechanics of the tasks.

\paragraph{\textbf{Steering}}
The 1D steering dataset was collected across two consecutive studies by \citet{cowley_frontiers_nodatE} and \citet{palomaki_link_2021}, in which 18 participants performed an obstacle-avoidance steering game. The game was designed to induce a state of flow by dynamically adjusting the task difficulty to match each participant’s skill level. Each participant completed 40 trials, resulting in an average of 124 minutes of gameplay per subject. Data were recorded at a sampling rate of 59 Hz using a Logitech G920 Driving Force steering wheel.

\paragraph{\textbf{Crank Rotation}}
The 1D crank-rotation dataset \citep{tessari_brownian_2024} captures one-dimensional circular movements, with 11 participants rotating a crank in either of two directions throughout each trial. They carried out multiple trials while maintaining one of three velocity levels: fast, comfortable, or slow. On average, each participant contributed 65 minutes of recorded data across 194 trials. The crank was mounted on a high-precision incremental optical encoder/interpolator set (Gurley Precision Instruments). During the task, participants’ elbows were supported by a canvas to minimize extraneous movement and keep the upper and lower arms in the plane of the crank. Crank angle was recorded at 200 Hz.

\paragraph{\textbf{Fitts's Task}}
20 participants performed a modified version of a 2D Fitts's task \citep{ilestz_ilestzfittslaw_msi_2024}. Using a computer mouse, participants moved a cursor to a target zone of varying size, with smaller zones representing higher task complexity. The dataset also includes segments where participants returned the cursor to the starting position. Data were recorded at 240 Hz, with each participant contributing an average of 34 minutes of movement data across 590 trials.

\paragraph{\textbf{Whack-A-Mole}}
The 2D Whack-A-Mole dataset \citep{meyer_mouse_2023} was collected via an online platform. A total of 363 participants each completed a single trial of a fast-paced, mouse-controlled game. Participants were instructed to click on a “mole” as quickly as possible when it appeared, and each trial lasted approximately 3 minutes. The data were recorded at an average sampling rate of 66 Hz. Although participants used various recording devices, only data from standard computer mice were included.

\paragraph{\textbf{Object Moving}}
In this 3D task, participants were instructed to move a physical cylinder from one position to another while avoiding an obstacle cylinder \citep{grimme_naturalistic_2012, raket_separating_nodate}. The obstacle varied in both position and length across trials. On average, each participant contributed 3 minutes of movement data across 160 trials. Movements were recorded using the Visualeyez (Phoenix Technologies Inc.) VZ 4000 motion capture system, with a wireless infrared light-emitting diode (IRED) attached to the object. Marker trajectories were recorded in three Cartesian dimensions at 110 Hz in a reference frame anchored to the table. The recorded signals were filtered with a low-pass filter at 5.5 Hz.

\paragraph{\textbf{Pointing}}
The 3D pointing dataset \citep{liao_characterizing_nodate} involves participants directing their movements toward a hollow quarter-sphere, which was repositioned by a robotic actuator at the start of each trial. Five participants contributed, on average, 11 minutes of data across 323 trials. Three-dimensional fingertip and target positions were sampled at 100 Hz using an optical tracking system (Optotrak 3020, Northern Digital Inc.), with IRED markers placed at both the fingertip and on the robotic actuator.

\paragraph{\textbf{Handwriting}}
The 3D handwriting dataset was compiled from two studies by \citet{handwriting1} and \citet{handwriting2}, in which 91 participants wrote words or sentences based on a given prompt. Participants contributed an average of 6 minutes of writing data, encompassing 79 recorded samples. A Wacom-enabled Samsung S7 FE tablet was used to record coordinates at a sampling rate of 370 Hz. Although most of the writing data is confined to a two-dimensional plane, the dataset also includes vertical displacement for midair strokes within 7 mm of the screen; these strokes are shown in light blue in Figure~\ref{fig:task_grouping}~(g).

These datasets span a variety of tasks, degrees of freedom, and sampling rates, providing a robust foundation for testing and validating a movement decomposition approach. For a more comprehensive description of the datasets, please refer to the respective original publications.

\begin{table}[ht!]
\centering
\renewcommand{\arraystretch}{1.2}
\caption{Summary of datasets grouped by degrees of freedom (\textit{D}).}
\label{tab:datasets}
\begin{tabular}{|l|l|l|l|l|l|l|}
\hline
\textbf{Dataset} &
\textbf{Size (h)} & \makecell{\textbf{SR}\\(\textit{s}\textsuperscript{-1})} & 
\makecell{\textbf{Avg.}\\\textbf{Dur. (s)}} & 
\makecell{\textbf{Num.}\\\textbf{Subjects}} &
\makecell{\textbf{Low Freq.}\\\textbf{Comp. R\textsuperscript{2}}} &
\textbf{License} \\
\hline

\multicolumn{7}{|c|}{\textbf{1D}} \\ \hline
Steering$^{a}$ & 37.2 & 59  & 186.1 & 18  & 99.2\% & CC BY 4.0 \\ \hline
Crank Rotation$^{b}$   & 11.9 & 200 & 20.1  & 11  & 99.6\% & CC BY 4.0 \\ \hline

\multicolumn{7}{|c|}{\textbf{2D}} \\ \hline
Fitts's Task$^{c}$   & 10.9 & 240 & 3.3   & 20  & 91.8\% & CC BY 4.0 \\ \hline
Whack-A-Mole$^{d}$   & 17.4 & 66  & 172.5 & 363 & 93.4\% & CC BY 4.0 \\ \hline

\multicolumn{7}{|c|}{\textbf{3D}} \\ \hline
Object Moving$^{e}$   & 0.5  & 110 & 1.1   & 10  & 99.8\% & CC BY 4.0 \\ \hline
Pointing$^{f}$ & 0.9  & 100 & 2     & 5   & 99.7\% & CC BY 4.0 \\ \hline
Handwriting$^{g}$   & 6.5  & 370 & 3.3   & 91  & 89.3\% & Custom$^*$ \\ \hline
\end{tabular}

\vspace{1ex}
\footnotesize
\textbf{Notes.} Table~\ref{tab:datasets} reports the total recording time in hours (\textbf{Size}), average sampling rate in samples s\textsuperscript{–1} (\textbf{SR}), average trial duration in seconds (\textbf{Avg.~Dur.}), number of participants (\textbf{Num.~Subjects}), and the proportion of variance explained by the low-frequency component of the signal (\textbf{Low Freq.~Comp.~R\textsuperscript{2}}).  See Section~\ref{sec:data_prep} for calculation details.  
\textit{CC BY 4.0} = Creative Commons Attribution 4.0 International License; $^*$custom licence permits research use only.

\textbf{Data repository references:}
$^{a}$\;\citet{data_cogcarsim};
$^{b}$\;\citet{jameshermus_jameshermusbrownianprocess_2023, noauthor_crank_data_nodate};
$^{c}$\;\citet{ilestz_ilestzfittslaw_msi_2024};
$^{d}$\;\citet{meyer_continuous_2021};
$^{e}$\;\citet{raket_larslaubochum_movement_data_2022};
$^{f}$\;\citet{noauthor_data_nodate};
$^{g}$\;\citet{handwriting_data}.
\end{table}

\subsection{Data Preparation}
\label{sec:data_prep}

In this section, we describe the preprocessing steps applied to ensure consistency across signals while preserving features necessary for accurate submovement decomposition. The main steps are the computation of \emph{Signed Tangential Velocity (STV)} and signal resampling.

\paragraph{\textbf{Signed Tangential Velocity.}}
Tangential velocity \(v_T\) quantifies the speed along an object's trajectory, calculated from discrete position samples \(\mathbf{r}_i\).
For the 3D case, \(\mathbf{r}_i = (x_i, y_i, z_i)\). This measure is widely used for identifying submovements \citep{spurious2, liao_characterizing_nodate, review_paper1}. The velocity vector is determined via finite differences:

\begin{equation}
\mathbf{v}_i = \frac{\mathbf{r}_i - \mathbf{r}_{i-1}}{\Delta t}, \quad \Delta t = t_i - t_{i-1},
\end{equation}

with the tangential velocity defined as its magnitude:


\begin{equation}
v_i^T = \|\mathbf{v}_i\| = \frac{|\mathbf{r}_i - \mathbf{r}_{i-1}|}{\Delta t} = \frac{\sqrt{\sum_{j=1}^{n} (r_i^j - r_{i-1}^j)^2}}{\Delta t},
\end{equation}

where \(r_i^j\) represents the \(j\)-th component of position vector \(\mathbf{r}_i\) in an \(n\)-dimensional space.

\begin{figure}[ht!]
    \centering
    \begin{subfigure}{0.46\textwidth}
        \includegraphics[width=\linewidth]{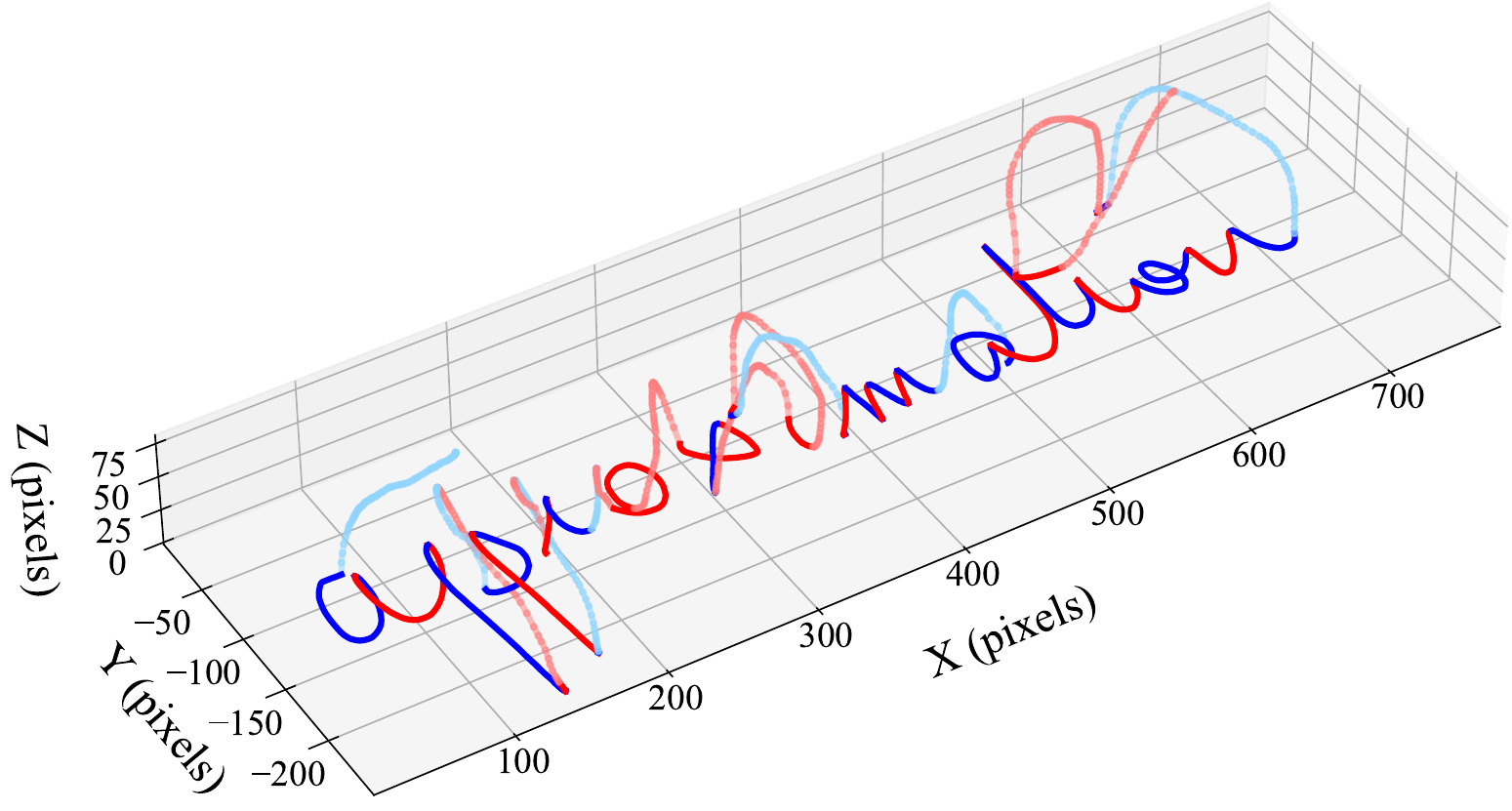}
        \caption{Writing path, colored by the \textbf{STV} sign.}
    \end{subfigure}
    \begin{subfigure}{0.42\textwidth}
        \includegraphics[width=\linewidth]{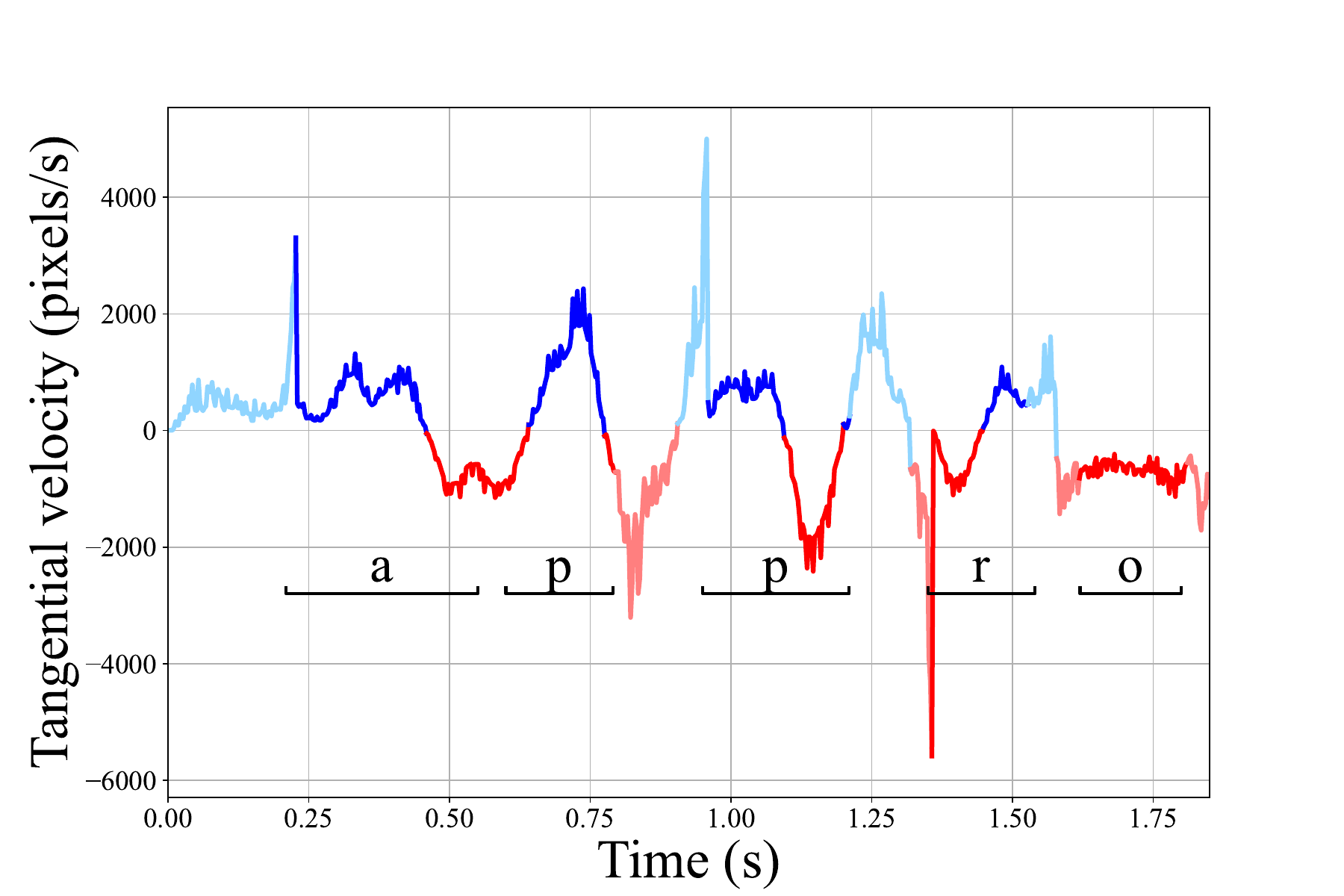}
        \caption{\textbf{STV} for the first five letters.}
    \end{subfigure}
    \caption{Visualization of the signed tangential velocity (\textbf{STV}). (a) The 3D plot shows the pen tip's path while writing the word ``approximation''. Light, translucent colors indicate midair hovering, whereas vivid, opaque colors denote writing on the tablet. Blue represents positive \textbf{STV} and red represents negative \textbf{STV}; color switches indicate significant and quick direction changes. (b) The corresponding \textbf{STV} for the first five letters. Brackets mark the segments associated with each letter. Note that while transitions often occur near the zero ordinate, this is not always the case due to sampling limitations at the exact zero crossing.}
    \label{fig:signed_velo}
\end{figure}

Complex movements often comprise overlapping submovements that either \emph{compound} (reinforce) or \emph{counteract} (oppose) each other. While this distinction is straightforward in one dimension (e.g., steering), multidimensional movements require a more nuanced approach that accounts for relative angles between submovements.
Standard tangential velocity (\(v_i^T\)) captures speed magnitude but lacks critical directional information. In ideal conditions, direction reversals would appear as momentary zeros in speed, aligning with the 2/3 Power Law \citep{zago_speed-curvature_2017}, but sensor limitations often obscure these subtle transitional cues. Figure \ref{fig:signed_velo} demonstrates how acute changes in movement direction coincide with lower absolute velocity values.

To address this limitation, we developed the \textit{Signed Tangential Velocity (STV)}, which preserves information about acute direction changes in movement paths. Unlike conventional tangential velocity, which is always positive, STV \(v_i^{T,\pm}\) incorporates sign changes that reflect directional reversals exceeding $90^\circ$.

To avoid capturing noise-induced reversals, we smooth the velocity measurements by averaging over \(n\) consecutive samples:

\begin{equation}
\overline{\mathbf{v}}_i = \frac{1}{n} \sum_{j=i}^{i+n-1} \mathbf{v}_j.
\end{equation}

The averaging window \(n\) is set to cover $\frac{1}{60}$ seconds to keep high temporal resolution, with a minimum of 3 samples to effectively reduce noise. We then compute the angle between successive smoothed velocity vectors:

\begin{equation}
\theta_i = \cos^{-1} \left( \frac{\overline{\mathbf{v}}_{i-n} \cdot \overline{\mathbf{v}}_i}{\|\overline{\mathbf{v}}_{i-n}\|\,\|\overline{\mathbf{v}}_i\|} \right).
\end{equation}

We identify acute direction changes - akin to generalized reversals in movement trajectory - when $\theta_i$ exceeds $90^\circ$ and represents a local maximum in the angle signal. The path parity at time $i$ is defined as:
\begin{equation}
    p_i=\text{number of directional reversals between $t=1$ and $t=i$}
\end{equation}


The STV is then defined by:

\begin{equation}
v_{i}^{T,\pm} = (-1)^{p_i}v_i^T.
\end{equation}

Each detected reversal flips the velocity sign, encoding the acute directional change and indicating when a current submovement counteracts the previous one. While sensor noise may cause false detection of directional switches during periods of little or no movement, such cases do not significantly impact submovement decomposition. When there is no overlap between submovements, a sign switch simply indicates the absence of intersecting submovements on either side of the reversal.

In summary, STV provides a more informative measure than the traditional tangential velocity by highlighting acute direction changes. This measure is particularly useful for motor tasks requiring frequent directional adjustments, such as handwriting \citep{fast_writting}, and facilitates the disentanglement of compound submovements. See Figure~\ref{fig:signed_velo} for an illustrative example.

\paragraph{\textbf{Resampling.}}  
All recorded signals were resampled to a uniform rate of 60 Hz, with missing positional data recovered through linear interpolation.

\paragraph{\textbf{No Additional Filtering.}}
We deliberately avoided additional filtering to evaluate model performance under realistic noise conditions and preserve the potential for real-time applications. In particular, we avoided commonly used non-causal filtering methods \citep{grimme_naturalistic_2012, liao_characterizing_nodate} that are unsuitable for online processing.

To estimate the upper limits of reconstructing the signals using smooth primitives, we analyzed the low-frequency components of the data. Specifically, we computed the coefficient of determination (\(R^2\)) for the low-frequency signal obtained after applying a low-pass filter at 10 Hz. This cutoff was chosen based on the upper limit of human-generated movement frequencies reported in prior studies \citep{review_paper1}.

In four datasets, frequencies above 10 Hz contributed less than 1\% of total variance. However, the Whack-A-Mole, Fitts's Task, and Handwriting datasets showed significantly higher proportions of variance unexplained by low-frequency components—93.4\%, 91.8\%, and 89.3\%, respectively—indicating substantial noise levels and establishing practical reconstruction boundaries. All numerical details are provided in Table \ref{tab:datasets}.

Notably, the Object Moving dataset was the only one explicitly mentioned to be pre-processed with a low-pass filter (cutoff frequency of 5.5 Hz), which is consistent with its low-frequency component accounting for 99.8\% of the total variance—the highest among all datasets.

\paragraph{\textbf{Scaling and Additive Noise.}}
In the final preprocessing step, the velocity signal was scaled to a root mean square of 1 before network input. Assuming the velocity signal in the limit has zero mean, this normalization corresponds to a standard deviation of 1.

To evaluate model robustness, we supplemented the clean signal with Gaussian noise at two levels, yielding signal-to-noise ratios (SNRs) of 20 dB and 10 dB, representing medium and high noise conditions respectively.

\section{Semi-Supervised Submovement Decomposition}


In this work, we introduce a novel semi-supervised approach to submovement decomposition that leverages both synthetic data and unlabeled real-world movement data. 

Our approach begins by creating synthetic training data with known submovement characteristics based on minimum-jerk velocity profiles, with parameters drawn from ranges reported in previous literature. This synthetic data, which contains ground-truth information about submovement timings, durations, and displacements, is used to pre-train a neural network that decomposes velocity time series into submovement components.

The trained model is then applied to unlabeled human movement data to obtain initial predictions of submovement sequences. From these predictions, we estimate the statistical distributions of key submovement parameters in the real human motion data. These learned distributions are used to generate updated synthetic data that more closely mimics real movements, which is then used to refine the model. This iterative process creates a feedback loop between model training and synthetic data generation that progressively improves decomposition accuracy.  

Our approach follows two main stages:
\begin{enumerate}
    \item \textbf{Initial Stage:} 
    \begin{itemize}
        \item \textbf{Synthetic Pre-training:} Generate an initial set of synthetic velocity profiles with known submovements and train a decomposition model on this data.
    \end{itemize}
    
    \item \textbf{Iterative Refinement Stage:} The following steps are repeated until convergence:
    \begin{itemize}
        \item \textbf{Pseudo-Labeling and Distribution Estimation:} The current model processes unlabeled real movement data to infer submovement labels (pseudo-labels). These pseudo-labels are analyzed to approximate the statistical distributions of submovement parameters present in the real data.
        \item \textbf{Synthetic Data Refinement:} A new synthetic dataset is generated by sampling submovement parameters from the newly estimated distributions.
        \item \textbf{Model Re-training:} The model is further trained using a combination of the refined synthetic data and the original cold-start synthetic data, incorporating updated statistics.
    \end{itemize}
\end{enumerate}

\subsection{Temporal Fully-Convolutional Detector}
\label{sec:detector}

\begin{figure}[ht]
    \centering
    \includegraphics[width=\textwidth]{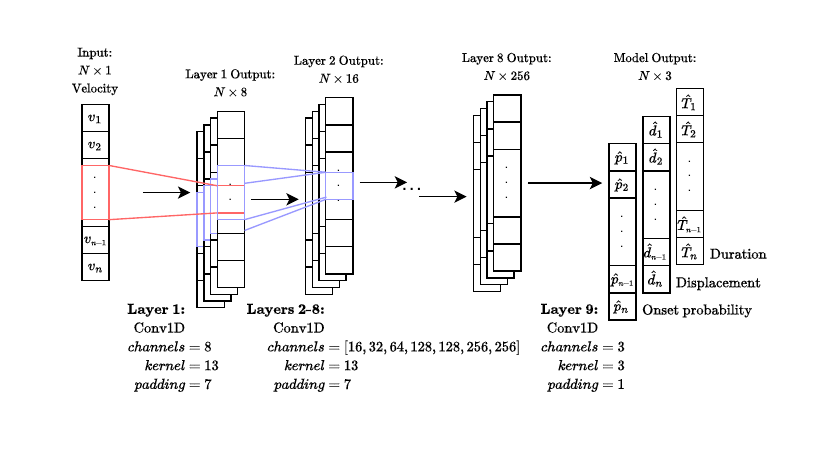}
    \caption{Architecture of the Temporal Fully-Convolutional Network (TFCN) Detector module. The network processes a 1D velocity time series of length $N$ through eight hidden convolutional layers with increasing channel dimensions $[8, 16, 32, 64, 128, 128, 256, 256]$. The red lines schematically illustrate the receptive field of a single first-layer neuron (kernel size = 13), while blue lines demonstrate how the second layer's neurons process the 8-dimensional outputs from the first layer. Most layers use kernel size = 13 with padding = 7, creating a total receptive field of 99 time steps. Batch normalization and dropout (0.2) are applied between layers. For each time step, the model outputs three parallel predictions: onset probability $\hat{P}_t$, submovement duration $\hat{T}_t$, and displacement $\hat{d}_t$, maintaining the original sequence length $N$.}
    \label{fig:detector}
\end{figure}

The Detector module, which forms the backbone of our submovement decomposition method, is implemented as a Temporal Fully-Convolutional Network (TFCN) \citep{tcnn}.
This architecture is a specialized 1-dimensional convolutional neural network (1D CNN) well-suited for sequence-to-sequence processing of time-series data.

While TFCN models offer less contextual capacity than recurrent neural networks (RNNs) and transformers, they provide significant advantages through high parallelization and linear time and memory complexity ($O(N)$), compared to the sequential nature of RNNs and the quadratic complexity ($O(N^2)$) of standard Transformer architectures. This architecture is particularly effective for tasks where local temporal features are crucial, such as speaker identification \citep{snyder_x-vectors_2018} and EEG signal classification \citep{eeg_tcnn}.

The network processes a 1-dimensional velocity time series of length $N$ and produces three parallel output sequences of the same length:

\begin{itemize}
    \item Onset Probability ($\hat{p}$): The estimated probability ($\hat{p}_t \in [0, 1]$) that a submovement initiates at time step $t$.
    \item Duration ($\hat{T}$): The predicted duration ($\hat{T}_t$) for a potential submovement starting at time step $t$.
    \item Displacement ($\hat{d}$): The predicted net displacement ($\hat{d}_t$) associated with a potential submovement starting at time step $t$.
\end{itemize}

Our TFCN architecture consists of 9 convolutional layers. To capture contextual information, we use larger kernels for hidden layers. Specifically, most layers employ a kernel size of 13. This configuration provides each output prediction with a receptive field spanning 99 input time steps, corresponding to approximately 1.65 seconds of temporal context. Both the kernel size and receptive field extent were determined through hyperparameter optimization.

The complete architecture contains approximately 1.6 million learnable parameters. Figure~\ref{fig:detector} presents a schematic representation of the Detector architecture.


\subsection{Partially Differentiable Reconstructor}
\label{sec:reconstructor}

\begin{figure}[ht]
    \centering
    \includegraphics[width=\textwidth]{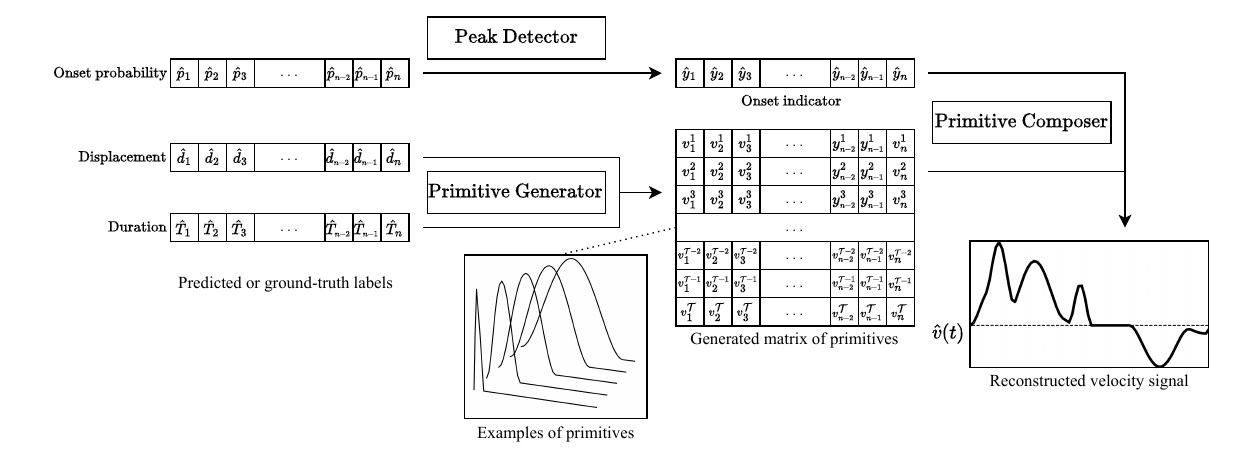}
    \caption{Architecture of the Partially Differentiable Reconstructor module. The pipeline processes three input sequences derived from either ground-truth labels or Detector predictions: onset probability, displacement, and duration. The Peak Detector transforms continuous onset probabilities ($\hat{p}_1, \hat{p}_2, ..., \hat{p}_n$) into binary onset indicators ($\hat{y}_1, \hat{y}_2, ..., \hat{y}_n$). For each detected onset, the Primitive Generator creates a minimum-jerk velocity profile using the corresponding displacement ($\hat{d}_t$) and duration ($\hat{T}_t$) values, producing a matrix of primitives where each row ($v^i_1, v^i_2, ..., v^i_n$) represents a single time step across all generated submovements, and each column represents a complete submovement profile. The Primitive Composer superimposes these profiles through temporal summation to produce the final reconstructed velocity signal $\hat{v}(t)$. The bottom-left inset shows examples of individual minimum-jerk primitives with varying displacements and durations. The entire process maintains differentiability with respect to displacement and duration parameters, enabling gradient-based optimization during training.}
    \label{fig:reconstructor}
\end{figure}

The Reconstructor module enables the Detector model to learn decompositions that lead to better reconstructions. It transforms discrete submovement parameters into continuous velocity signals, enabling end-to-end training through differentiable signal reconstruction. Figure~\ref{fig:reconstructor} illustrates the reconstruction pipeline.

The module processes submovement labels that can originate from two sources:
\begin{itemize}
    \item Synthetic Ground-truth: Sparse labels generated during synthetic data creation, consisting of binary onset indicators and precise values for displacement and duration.
    \item Detector Predictions: Continuous sequences of onset probabilities ($\hat{p}_t$), durations ($\hat{T}_t$), and displacements ($\hat{d}_t$) produced by the Detector module for each time step $t$.
\end{itemize}

When processing Detector predictions, the onset probability sequence first passes through the Peak Detector that identifies local maxima exceeding a predetermined threshold ($p \ge 0.5$). These peak locations represent the estimated onset times of individual submovements. For each identified onset (whether from ground-truth or prediction), the Primitive Generator creates a minimum-jerk velocity profile using the corresponding displacement and duration values. This generates a matrix of primitives, where each column represents a single submovement's velocity profile across all time steps.

The Primitive Composer then temporally superimposes (sums) these individual profiles to synthesize the final reconstructed velocity signal ($\hat{v}(t)$). While we implement minimum-jerk profiles as our standard submovement model based on motor control literature, our framework's modular design accommodates alternative profile functions.

A key feature of the Reconstructor is its partial differentiability with respect to input parameters. Specifically, gradients can propagate backward from the reconstruction loss (comparing $\hat{v}(t)$ with the target signal) through the superposition and profile generation steps to the duration ($\hat{T}_t$) and displacement ($\hat{d}_t$) inputs. This gradient flow enables the optimization process to directly minimize signal reconstruction error by adjusting the predicted parameters, creating a biologically plausible decomposition that faithfully reproduces the original movement dynamics.


\subsection{Synthetic Data Generation}
\label{sec:synthetic_data}

For initial model pre-training, we generate cold-start synthetic data using weakly informed prior distributions of submovement parameters. Each synthetic sample consists of a velocity signal $v \in \mathbb{R}^N$ and its corresponding ground-truth labels $y \in \mathbb{R}^{3 \times N}$, constructed through the following process:

Submovement parameters are sampled from biologically plausible distributions informed by motor control literature. Durations ($T$) are uniformly sampled between 85 ms and 1000 ms (approximately 5 to 60 samples), capturing the typical temporal scales of human submovements. Displacement values ($d$) are scaled proportionally to duration and modulated by a polynomial-weighted Gaussian random variable, using $|x|$ as the underlying weighting function. This approach reduces the relative frequency of submovements with lower peak velocities and yields signals with approximately unit standard deviation.

The temporal structure of submovements is controlled by sampling inter-submovement intervals as a fraction of the preceding submovement's duration, drawn uniformly from the range 0 to 1.5. This allows for temporal patterns ranging from extreme overlap to widely spaced submovements. We enforce a minimum separation of 2 samples ($\approx 33 \text{ ms}$) between adjacent onsets to prevent temporal aliasing effects and ensure distinct submovements. Each synthetic trial's total duration varies between 100 and 500 samples (approximately 1.5 to 8 seconds), providing sufficient variability in sequence length to simulate realistic movement patterns.

To enhance model robustness, we apply Gaussian noise ($\eta \sim \mathcal{N}$) to the reconstructed velocity signals, simulating measurement noise and biological variability, present in real movement data. This augmentation helps prevent overfitting and improves generalization to real-world signals.

The ground-truth labels $y$ are structured as a $3 \times N$ matrix, where each column represents a time step $t$, with rows corresponding to:
\begin{enumerate}
    \item Onset Indicator ($y_t$): Binary values where 1 indicates the initiation of a submovement at time $t$, and 0 otherwise.
    \item Duration ($T_t$): The duration parameter of the submovement beginning at time $t$ (zero for non-onset time steps).
    \item Displacement ($d_t$): The total displacement associated with the submovement starting at time $t$ (zero for non-onset time steps).
\end{enumerate}

\begin{figure}[ht]
    \centering
    \includegraphics[width=\textwidth]{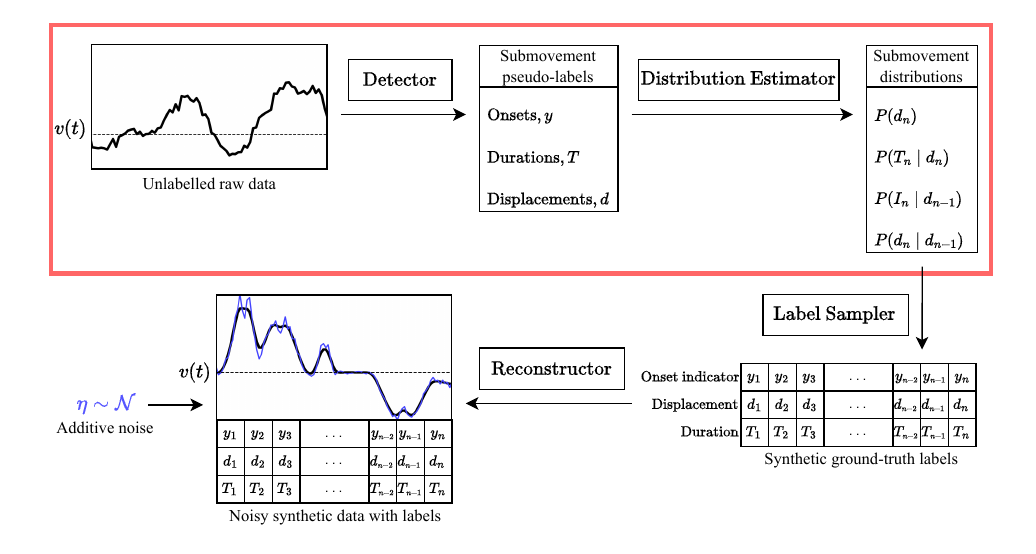}
    \caption{Semi-supervised data generation pipeline. The upper section (red box) illustrates the distribution estimation process: unlabeled raw velocity data $v(t)$ is processed by the Detector to generate pseudo-labels for onsets, durations, and displacements. These labels feed into the Distribution Estimator, which calculates probability distributions for various submovement parameters, including marginal displacement distributions $P(d_n)$ and conditional distributions: duration given displacement $P(T_n | d_n)$, onset-to-onset interval given displacement $P(I_n | d_n)$, and next displacement given previous one $P(d_n | d_{n-1})$.
    The lower portion illustrates how synthetic training data is created: the Label Sampler draws parameter values to create synthetic ground-truth labels, which the Reconstructor transforms into velocity signals. Additive Gaussian noise $\eta \sim \mathcal{N}$ creates realistic noisy synthetic data with known ground-truth labels for model training.
    }
    \label{fig:bootstrap}
\end{figure}

This labeled synthetic dataset enables supervised pre-training of our detection model before applying it to unlabeled real movement data. The process of generating synthetic data is illustrated in the lower part of Figure~\ref{fig:bootstrap}.

\subsection{Bootstrapped Distribution Refinement}
\label{sec:bootstrap}

After initial pre-training on cold-start synthetic data, we implement an iterative refinement process to progressively align our synthetic data generation with the statistical properties of real human movements. This bootstrap procedure creates a positive feedback loop between model performance and training data quality.

The process begins by applying the pre-trained model to unlabeled human motion data, generating pseudo-labels that represent the model's current best estimate of submovement decomposition. From these predictions, we extract and estimate the statistical distributions of key submovement characteristics including displacement, duration, and temporal relationships between consecutive submovements.

For distribution modeling, we employ non-parametric conditional multivariate kernel density estimation (KDE) through the `fastkde` package \citep{kde}. This approach allows us to capture complex, potentially multimodal distributions without imposing restrictive parametric assumptions. To balance computational efficiency with statistical accuracy, we model the following key relationships:

\begin{itemize}
    \item Marginal distribution of displacement: $P(d_n)$
    \item Duration conditional on displacement: $P(T_n | d_n)$
    \item Onset-to-onset interval conditional on displacement: $P(I_n | d_n)$
    \item Sequential displacement dependency: $P(d_n | d_{n-1})$
\end{itemize}

KDE becomes computationally intensive with an increased number of dimensions. Thus this factorization of the joint distribution focuses computational resources on modeling the most informative relationships while maintaining tractability during training. We use displacement as our primary conditioning variable because it contains the most information about submovement characteristics and strongly influences the relationships between consecutive submovements (whether compounding or counteracting, primary or secondary).
Our goal here is not perfect statistical modeling of organic data but rather to generate synthetic data that mimics real movements closely enough while maintaining beneficial variance to improve model robustness.

During training, we regenerate synthetic data after each epoch using the updated distribution estimates. This creates a progressive refinement cycle where:
\begin{enumerate}
    \item The model's prior shifts toward labels that can generate data mimicking organic movements.
    \item The model's ability to discriminate between labels that produce similar but distinct velocity signals increases, as it is exposed to more training samples with labels that result in velocity signals closely resembling those in organic data.
\end{enumerate}

This convergence is evidenced by improved reconstruction performance.

To prevent overfitting to potentially imperfect distribution estimates, we maintain diversity in our training data by mixing distribution-based synthetic samples (50\%) with original cold-start synthetic data (50\%). This mixture helps preserve model robustness while still guiding it toward more biologically plausible decompositions.

\subsection{Training Procedure and Loss Functions}

All models were implemented using the PyTorch framework \citep{pytorch} and trained in a Google Colab Pro environment equipped with an NVIDIA L4 GPU. Training was conducted using a batch size of 512, with each epoch comprising 1,000 batches (512,000 samples in total). For each batch, the duration was sampled between 100 and 500 samples (approximately 1.5 to 8 seconds). Batch normalization was employed throughout the network to stabilize the training.

\paragraph{\textbf{Pre-Training Phase.}} The training process began with a base model fitted exclusively on synthetic labels generated via cold-start heuristics. This pre-training phase extended for 25 epochs. To prevent disruption of early training dynamics, the reconstruction objective was integrated only after the 10\textsuperscript{th} epoch.
A dropout rate of 0.2 was applied during the first 20 epochs, and subsequently disabled to avoid negatively affecting the regression heads. The pre-training phase typically required approximately 2 hours of computation time.

\paragraph{\textbf{Fine-Tuning Phase.}}
After pre-training, the base model was fine-tuned across multiple human motion datasets.
Each epoch began with statistical estimation from the training splits of human-motion datasets, which informed the generation of updated synthetic signals. Batches were balanced with 50\% weakly-informed synthetic data and 50\% signals sampled equally from each human motion dataset statistics. Dropout remained disabled during this stage. The learning rate decayed exponentially throughout training, starting at $1\times 10^{-3}$ and gradually decreasing to $1\times 10^{-5}$. The fine-tuning typically required approximately 1.5 hours.

\paragraph{\textbf{Loss Function and Normalization Strategy.}}
The model's optimization objective comprised four components: binary cross-entropy (BCE) for submovement onset detection, and mean squared error (MSE) losses for predicting duration, predicting displacement, and reconstructing the signal. To address the significant class imbalance between onset (positive) and non-onset (negative) samples, the BCE loss was weighted to prevent domination by the majority class:

\[
\mathcal{L}_{\text{BCE}}(y, \hat{p}) = -\frac{1}{N} \sum_{t=1}^{N} \left[ \alpha y_t \log(p_t) + (1 - \alpha)(1 - y_t) \log(1 - p_t) \right],
\]

where $\alpha$ represents the positive-class weighting factor, $y_t$ denotes the ground-truth onset label, and $p_t$ is the predicted probability at time step $t$.

During the fine-tuning, there is a risk that the training could enter a loop of reinforcing decomposition into an increasingly larger number of submovements. While this would lead to higher reconstruction accuracy, it would ultimately result in implausible decompositions. To prevent this, during the fine-tuning phase, we applied an adaptive weighting scheme:

\[
\mathcal{L}_{\text{BCE}}(y, \hat{p})^{\text{fine}} = -\frac{1}{N} \sum_{t=1}^{N} \left[ \alpha \sqrt{|d_t|} \cdot y_t \log(p_t) + (1 - \alpha) \cdot \beta_t \cdot \gamma \cdot (1 - y_t) \log(1 - p_t) \right],
\]

where:
\begin{align*}
  \gamma &= \max\!\Bigl(1, \frac{|\hat{\mathcal{O}}|}{|\mathcal{O}|}\Bigr)
    & 
    \beta_t &= \begin{cases}
    0.25, & \text{if } \min_{s \in \mathcal{O}} |t-s| = 1 \\
    0.5, & \text{if } \min_{s \in \mathcal{O}} |t-s| = 2 \\
    1, & \text{if } \min_{s \in \mathcal{O}} |t-s| \geq 3
    \end{cases}
\end{align*}

Here, $\sqrt{|d_t|}$ weights the positive class loss by the square root of the absolute displacement of ground-truth submovements, prioritizing detection of larger movements. The ratio $\gamma$ penalizes the model for producing more false positives by comparing the number of detected submovement onsets $|\hat{\mathcal{O}}|$ to the number of ground-truth submovement onsets $|\mathcal{O}|$. Additionally, $\beta_t$ applies an exponential decay based on the distance to the nearest ground-truth submovement onset, making false positives further from actual submovements more costly.

Duration and displacement losses are computed over the set of ground-truth submovement onset timestamps $\mathcal{O}$:
\[
\mathcal{L}_{\text{MSE}}(T, \hat{T}) = \frac{1}{|\mathcal{O}|} \sum_{t \in \mathcal{O}} \left( \hat{T}_t - T_t \right)^2,
\]

\[
\mathcal{L}_{\text{MSE}}(d, \hat{d}) = \frac{1}{|\mathcal{O}|} \sum_{t \in \mathcal{O}} \left( \hat{d}_t - d_t \right)^2,
\]
where $\hat{T}_t$ and $\hat{d}_t$ represent the predicted duration and displacement at onset time $t$, while $T_t$ and $d_t$ denote the corresponding ground-truth values.

Signal reconstruction loss is defined over the complete time sequence using the clean (noise-free) signal as the target:

\[
\mathcal{L}_{\text{MSE}}(v, \hat{v}) = \frac{1}{N} \sum_{t=1}^{N} \left( v_t - \hat{v}_t \right)^2,
\]

where $v_t$ is the true clean signal and $\hat{v}_t$ its reconstruction.

To ensure balanced contribution from heterogeneous loss terms, each loss component is normalized by its mean value during the current epoch, and scaled by the mean of the detection loss:

\[
\mathcal{L}_{\text{total}} = \sum_{i=1}^{4} \tilde{w}_i \cdot \frac{\mathcal{L}_i}{\mathbb{E}[\mathcal{L}_i]} \cdot \mathbb{E}[\mathcal{L}_{\text{BCE}}],
\]

where $\tilde{w}_i$ are raw loss weights, and $\mathbb{E}[\cdot]$ denotes the epoch mean of the corresponding loss. This dynamic normalization ensures that all components contribute meaningfully to the total objective, while maintaining interpretability throughout training. Since reconstruction optimization begins only in epoch 10 of the pre-training phase, its corresponding weight $\tilde{w}_4$ is initially set to zero.

\begin{figure}[ht]
    \centering
    \includegraphics[width=\textwidth]{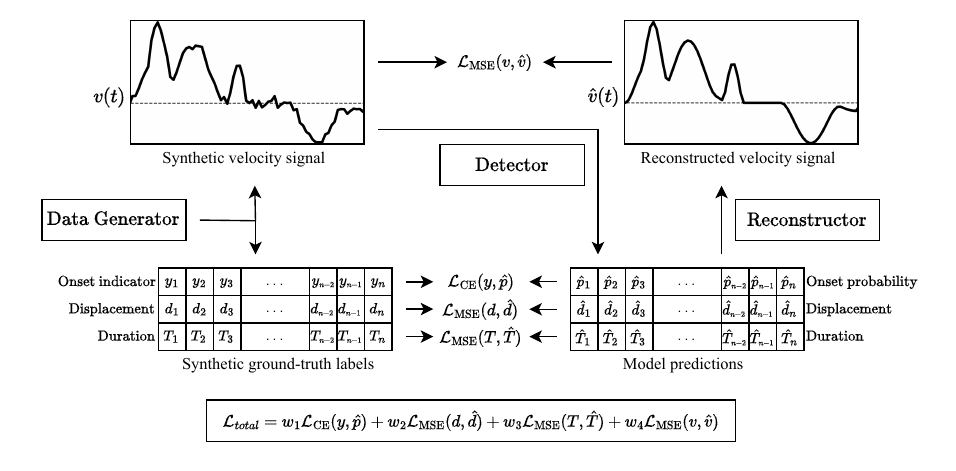}
    \caption{Comprehensive illustration of the training process for \ssumo model. The Data Generator produces synthetic velocity signals (left, top) with associated ground-truth labels for onset indicators, displacements, and durations (left, bottom). Noisy synthetic signals are fed into the Detector model, which outputs both predicted submovement characteristics and a reconstructed velocity signal (right). Four distinct loss components are calculated: (1) cross-entropy loss ($\mathcal{L}_{\text{CE}}$) between true and predicted onset probabilities, (2-3) mean squared error losses ($\mathcal{L}_{\text{MSE}}$) for displacement and duration predictions, and (4) reconstruction loss ($\mathcal{L}_{\text{MSE}}$) between the clean input signal and model reconstruction. These components are weighted and combined into a total loss function ($\mathcal{L}_{\text{total}}$) that guides model optimization. Note that while training inputs may contain noise, the reconstruction target is the clean signal, enabling the model to learn denoising capabilities.}
    \label{fig:training}
\end{figure}

\section{Experiments}

\subsection{Evaluation Metrics}
\label{sec:metrics}

Model performance is measured with complementary \emph{supervised} and \emph{unsupervised} criteria, selected according to whether ground-truth submovement annotations are available.

\paragraph{\textbf{Supervised Metrics.}}
Synthetic trials provide exact onset, duration and displacement labels, allowing us to evaluate both \textit{event detection} and \textit{parameter regression}. To assess onset detection accuracy, we compute the $F_1$ score. A predicted onset is counted as a true positive if it occurs within five samples ($\approx$83 ms) of a ground-truth event, matches the displacement sign, and is the closest such prediction to that ground truth onset. The distance threshold is based on the reported minimum interval between submovement onsets \citep{high_freq1}.

Predictions without a corresponding match are considered false positives, while unmatched ground-truth events contribute to the false negative count.
For each ground-truth onset we take the closest prediction and compare the associated duration and displacement.  Goodness of fit is reported with the coefficient of determination, $R^{2}$.  Negative $R^{2}$ values—indicating fits worse than a constant predictor—are clipped to zero so that a small number of gross failures cannot dominate the average. In addition to the primary supervised metrics, we also report number of correctly detected and spurious submovements per second. All supervised scores are computed \emph{per recording} and then averaged. We additionally report the interquantile ranges (IQR) in selected cases to provide additional clarity.

\paragraph{\textbf{Unsupervised Metrics.}}
When only the velocity trace is known, fidelity is gauged by the $R^{2}$ between the model's reconstruction and the clean (pre-noise) input.  As in the supervised setting, $R^{2}$ is truncated at zero.  We also report the algorithm's predicted number of submovements per second, which serves as a proxy for over- or under-segmentation when no labels are present. Scores are again averaged across recordings, with IQR supplied whenever it improves interpretability.

Together these metrics capture both discrete event-detection accuracy and overall signal-level reconstruction quality.

\subsection{Baselines}

In this subsection we outline the two baseline algorithms re-implemented for comparison, closely following the procedures in earlier studies.

\paragraph{\textbf{Peak Detector.}}
A simple linear-time peak-detector assumes that every prominent peak in the velocity signal marks a single submovement, mirroring the logic of \citet{peak_detector, markkula_control, autogain}.  
Our implementation (Algorithm~\ref{fig:peak_detector} in the Appendix) adds support for variable submovement durations.  
The signal is first smoothed with a Gaussian of width \(\sigma\); samples whose absolute value falls below a threshold \(\theta\) are neglected, and the remaining peaks are treated as submovement onsets.

The two hyper-parameters were tuned by a grid search with iterative refinement, maximising the geometric mean of reconstruction \(R^{2}\) and onset-detection \(F_{1}\) on the synthetic test set.  
The search stopped when the best–worst score difference within a cycle dropped below 0.001 and produced \(\sigma = 2.5\) (samples) and \(\theta = 0.0375\).  

The detector is extremely fast and usually yields a satisfactory reconstruction, even on unconventional signals.  Its drawbacks are (i) strong sensitivity to \(\sigma\) and \(\theta\) and (ii) an inability to separate multiple submovements that merge into a single peak.

\paragraph{\textbf{Scattershot Optimisation Algorithm}.}
As an accurate but slower reference we employ a Scattershot-style optimiser based on the open-source \texttt{submovements} Python package \citep{submovement_python} and SciPy's \citep{scipy} \texttt{minimize} function with the L-BFGS-B solver. Ten random restarts are used for each call.

To keep runtime manageable on long sequences, decomposition is performed in a 120-sample window (\(\approx\)2 s) sliding by 60 samples (1 s).  Only onsets detected in the first half of each window are retained; onsets in the overlapping half are re-estimated in the next step, using both previously fixed submovements and the new context.

Key parameters were tuned by grid search, again maximising the geometric mean of reconstruction \(R^{2}\) and onset \(F_{1}\). 
The final settings are \(patience = 1\)—the number of additional submovements allowed after the error stops improving—and \(error\_threshold = 0.15\), the mean absolute reconstruction error at which new submovement insertion stops.  


\subsection{Synthetic Evaluation Protocol}
\label{sec:methods_synthetic}

To quantify accuracy under controlled conditions we synthesise velocity traces as a series of minimum-jerk submovements.  
Unlike the training data—where each displacement was drawn from a polynomial-weighted Gaussian—test displacements are sampled uniformly from \([-1,\,1]\) and then scaled by the submovement duration.  
This shift in distribution creates a distinct domain that contains a higher number of low-velocity submovements, making decomposition harder and therefore providing a stringent robustness test.

The onset-to-onset interval between successive submovements, expressed as a fraction of the first submovement’s duration, is drawn uniformly from one of three mutually exclusive ranges and one pooled range: 1.0–1.5 (no overlap), 0.5–1.0 (medium overlap), 0.0–0.5 (high overlap), 0.0–1.5 (pooled range).

Each synthetic trial lasts 16.7 s (1000 samples).  
For every overlap condition we generate 512 random trials.

Additive Gaussian noise is injected at three SNR levels:  
\(\infty\) dB (noise-free), 20 dB, and 10 dB.  
For the pooled overlap condition we perform a finer sweep from 50 dB down to 10 dB in 1 dB steps (plus \(\infty\)) for our method and the Peak Detector.  
Scattershot is evaluated only at \(\infty\), 20 dB, and 10 dB because of its high runtime.

To isolate the contribution of key design choices we perform ablation studies--comparing the full model with four variants, all trained and tested on the same synthetic benchmark:

\begin{itemize}
  \item No Reconstruction Loss — the reconstruction loss term is removed during training;
  \item No Signed Tangential Velocity — both input velocity signal and displacements are rectified (sign removed);
  \item Shorter Receptive Field — receptive field reduced from 1.65 s to 0.85 s;
  \item No Noise Augmentation — the model is trained without additive noise.
\end{itemize}

Ablations are benchmarked on the pooled overlap condition at 20 dB SNR.

While evaluating against the baselines, the results use the model fine-tuned on the complete set of human-motion data.
This setting tests whether a model tuned on real data can still decompose ground-truth synthetic traces.
For ablation comparisons we use the corresponding \emph{non-fine-tuned} checkpoints so that only the targeted modification differs.

\subsection{Human-Motion Evaluation Protocol}

We further evaluate on the test splits of the seven human-motion datasets.
Because Scattershot is computationally expensive, we randomly select a 20 s segment from every recording in the Steering and Whack-A-Mole datasets.

When comparing with the baselines, the proposed method performance is reported under leave-one-dataset-out fine-tuning: each run fine-tunes on six datasets and tests on the hold-out set, thereby measuring generalisation to unseen task types.

To quantify the benefit of fine-tuning on real motion data, we evaluate three model variants: (i) the base network trained only on synthetic traces, (ii) an \emph{in-domain} version fine-tuned on the combined training sets of all seven datasets, and (iii) an \emph{out-of-domain} version fine-tuned on the six datasets that exclude the one used for testing (the same variant used in benchmark comparisons).

We also benchmark execution speed on human-motion recordings. Each method is run on ten randomly chosen segments from every dataset, and we report the mean time required to process one second of signal. CPU timings are obtained on a 2022 MacBook Pro with an Apple M2 processor and 16 GB of RAM; for the proposed model, we additionally measure batched inference on an NVIDIA L4 GPU in Google Colab.

\section{Results}

Having defined the evaluation metrics (Section \ref{sec:metrics}), 
the synthetic benchmark (Section \ref{sec:methods_synthetic}), 
and the two baseline algorithms, we now turn to the empirical findings.  
We first report performance on the synthetic traces because they provide
complete ground-truth labels and controlled overlap/noise conditions, 
allowing us to isolate algorithmic behaviour.  
Results on real movement data follow in Section \ref{sec:results_human}.

\begin{table}[ht]
    \centering
    \scriptsize
    \setlength{\tabcolsep}{2.8pt}
    \fontsize{8}{12}\selectfont 
    \caption{Comparison of the proposed method against the benchmarks on synthetic data. Shown are the average Coefficient of Determination (\(R^2\)) between true and detected kinematics—Velocity Signal Reconstruction (\textbf{Rec.}), Displacement (\textbf{Disp.}) and Duration (\textbf{Dur.})—and the onset‐detection \(F_1\)-score (\textbf{On.}). Results are reported under four overlap distributions (onset-to-onset distance as a fraction of the first submovement’s duration) and three additive-noise levels: no noise (\(\infty\)), SNR 20 dB and SNR 10 dB. Within each overlap–noise condition, the highest values are highlighted in \textbf{bold}, and the second-highest in \textit{italic}. The proposed method consistently yields the best performance across all conditions and almost all metrics.}
    \begin{tabular}{ll|cccc|cccc|cccc}
        \toprule
        \multirow{2}{*}{\parbox{1cm}{\centering \textbf{Onset-to-Onset}}} &        
        \multirow{2}{*}{\parbox{0.8cm}{\centering \textbf{SNR (dB)}}} &       
        \multicolumn{4}{c|}{\textbf{\ssumo}} & 
        \multicolumn{4}{c|}{\textbf{Peak Detector}} & 
        \multicolumn{4}{c}{\textbf{Scattershot}} \\
        \cmidrule(lr){3-6} \cmidrule(lr){7-10} \cmidrule(lr){11-14}
        & & 
        \textbf{Rec.} & \textbf{On.} & \textbf{Disp.} & \textbf{Dur.} & 
        \textbf{Rec.} & \textbf{On.} & \textbf{Disp.} & \textbf{Dur.} & 
        \textbf{Rec.} & \textbf{On.} & \textbf{Disp.} & \textbf{Dur.} \\

        \multirow{3}{*}{1--1.5}
         & $\infty$ & \textbf{1.00}       & \textbf{.988}        & \textbf{.998}        & \textbf{.990}        & \textit{.978}        & \textit{.925}        & \textit{.953}        & \textit{.815}        & .973                 & .661                 & .506                 & .292                 \\
         & 20    & \textbf{.998}        & \textbf{.941}        & \textbf{.995}        & \textbf{.944}        & .973                 & \textit{.837}        & \textit{.962}        & \textit{.571}        & \textit{.989}        & .621                 & .524                 & .324                 \\
         & 10    & \textbf{.989}        & \textbf{.851}        & \textbf{.986}        & \textbf{.879}        & .926                 & \textit{.547}        & \textit{.868}        & \textit{.066}        & \textit{.961}        & .409                 & .453                 & .064                \\
         \midrule

        \multirow{3}{*}{0.5--1}
         & $\infty$ & \textbf{.998}        & \textbf{.966}        & \textbf{.992}        & \textbf{.969}        & .937                 & \textit{.694}        & \textit{.817}        & \textit{.386}        & \textit{.973}        & .665                 & .611                 & .207                 \\
         & 20    & \textbf{.994}        & \textbf{.841}        & \textbf{.962}        & \textbf{.878}        & .929                 & \textit{.662}        & \textit{.829}        & \textit{.204}        & \textit{.983}        & .646                 & .622                 & .196                 \\
         & 10    & \textbf{.972}        & \textbf{.738}        & \textbf{.937}        & \textbf{.774}        & .869                 & \textit{.569}        & \textit{.749}        & .006        & \textit{.957}        & .502                 & .469                 & \textit{.019}                \\
         \midrule

        \multirow{3}{*}{0--0.5}
         & $\infty$ & \textbf{.985}        & \textbf{.792}        & \textbf{.705}        & \textbf{.658}        & .881                 & .370                 & .000                 & .000                 & \textit{.971}        & \textit{.494}        & \textit{.135}        & \textit{.008}        \\
         & 20    & \textbf{.983}        & \textbf{.679}        & \textbf{.540}        & \textbf{.398}        & .874                 & .374                 & .001                 & .000                 & \textit{.980}        & \textit{.512}        & \textit{.153}        & \textit{.006}        \\
         & 10    & \textit{.949}        & \textbf{.537}        & \textbf{.412}        & \textbf{.164}      & .832        & .406        & .001        & .000        & \textbf{.954}        & \textit{.528}        & \textit{.084}        & \textit{.000}               \\
         \midrule

        \multirow{3}{*}{\textbf{0--1.5}}
         & $\infty$ & \textbf{.996}        & \textbf{.934}        & \textbf{.951}        & \textbf{.919}        & .938                 & \textit{.647}        & .414                 & \textit{.250}        & \textit{.975}        & .644                 & \textit{.503}        & .152                 \\
         & 20    & \textbf{.993}        & \textbf{.821}        & \textbf{.878}        & \textbf{.794}        & .934                 & .608                 & .430                 & .133                 & \textit{.984}        & \textit{.626}        & \textit{.515}        & \textit{.155}        \\
         & 10    & \textbf{.974}        & \textbf{.707}        & \textbf{.813}        & \textbf{.655}        & .883                 & \textit{.506}        & .387                 & .004                 & \textit{.958}        & .479                 & \textit{.399}        & \textit{.011}       \\
        \bottomrule
    \end{tabular}
    \label{tab:synthetic}
\end{table}

\subsection{Synthetic Evaluation}

On synthetic data, our method outperforms both the Peak Detector and the Scattershot baseline across all onset-to-onset interval distributions and SNR levels for every metric, as it is demonstrated in Table \ref{tab:synthetic}. The sole exception occurs in the high‐overlap, high‐noise condition (0.0–0.5× at 10 dB SNR), where Scattershot achieves a velocity reconstruction \(R^2 = 0.954\) compared to our \(R^2 = 0.949\).

Overall, all methods yield acceptable velocity reconstruction (\(R^2>0.83\)), with Scattershot generally outperforming the Peak Detector. Both the proposed method and Scattershot maintain \(R^2 \ge 0.949\) across conditions. However, high velocity reconstruction alone does not guarantee accurate recovery of submovement parameters. Our proposed algorithm consistently delivers the most precise onset detection, displacement estimates, and duration estimates. In contrast, the baselines only exceed an onset-detection \(F_1\) of 0.70 under non-overlapping conditions for the Peak Detector. Our method’s \(F_1\) drops below 0.70 only in the high-overlap case at 20 dB and 10 dB SNR.

Duration reconstruction is particularly challenging for the baselines: both the Peak Detector and Scattershot drop below \(R^2=0.1\) in high-noise or high-overlap scenarios. The maximum duration \(R^2\) attained by Scattershot is 0.331, whereas our method never falls below \(R^2=0.65\) except in the most high-overlap scenario with additive noise. Displacement \(R^2\) follows a similar pattern: our method remains above 0.70 in all but the highest overlap and additive noise, while Scattershot peaks at 0.63 under medium overlap and low noise.

Under low- and medium-overlap regimes—where submovements remain partially distinguished by velocity peaks—the Peak Detector generally recovers kinematics more accurately than Scattershot. In high-overlap the Peak Detector’s displacement estimation, however, falls near zero, and for the combined 0.0–1.5× range, it never exceeds \(R^2=0.43\).

Interestingly, for both Peak Detector and Scattershot, onset-detection \(F_1\) increases with noise in the high-overlap condition. We attribute this to their tendency to decompose noisier signals into a larger number of submovements, which by chance improves matching between detected and true onsets. However, it's clear that all the method including ours fails to accurately decompose the high-overlap condition with the presence of high noise. 

\begin{figure}[ht]
  \centering
  \includegraphics[width=\textwidth]{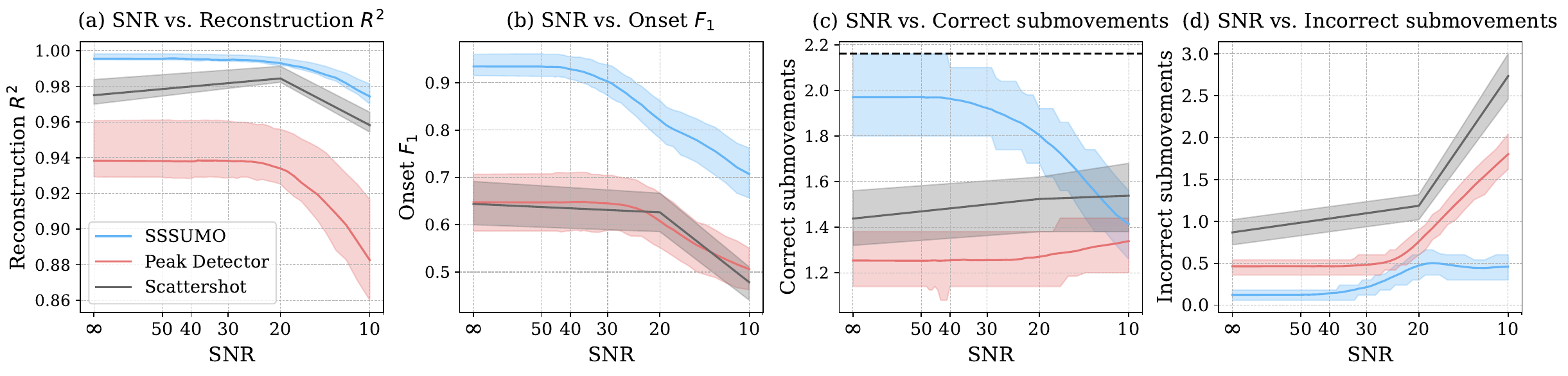}
  \caption{Dependence of performance on signal‐to‐noise ratio (SNR).  Solid lines show the mean and the shaded bands indicate the interquartile range for our method (blue), the Peak Detector (red) and Scattershot (gray) as SNR decreases from $\infty$ to 10~dB.  
    (a) Velocity‐reconstruction \(R^2\).  
    (b) Onset‐detection \(F_1\).  
    (c) Number of correctly detected submovements per second; the horizontal dashed line is the ground‐truth mean count.  
    (d) Number of incorrect submovements per second.  
  }
  \label{fig:metrics_snr}
\end{figure}

\paragraph{\textbf{Effect of Noise on Detection and Reconstruction.}}
Figure~\ref{fig:metrics_snr} examines how decreasing SNR impacts both velocity reconstruction and submovement detection. In panel (a), our method maintains velocity‐reconstruction \(R^2 > 0.99\) down to 17~dB, declining only to \(\approx 0.97\) at 10~dB. In contrast, both baselines start lower (\(R^2 \approx 0.94\)–0.98) even at high SNR. The slight performance bump for Scattershot at moderate noise likely arises because the optimizer runs longer in noisier signals. Nevertheless, Scattershot underperforms our method at every noise level considered.

Panel (b) shows onset‐detection \(F_1\): our algorithm stays above 0.90 until 30~dB and ends near 0.71 at 10~dB, while Peak Detector and Scattershot begin around 0.64–0.65 and fall to \(\sim 0.50\). Notably, despite better reconstruction, Scattershot does not improve \(F_1\) relative to Peak Detector.

Panels (c) and (d) break detection performance into true and false submovement counts per second. Panel (c) illustrates that no method recovers every true submovement: our approach achieves mean recall above 0.90 down to 34~dB, with the upper IQR reaching 100\% above 40~dB; recall drops to 0.65 at 10~dB. Interestingly, the baselines increase recall with noise, rising from 0.55–0.66 to 0.62–0.71 at 10~dB. Panel (d) explains this gain: false‐positive detections for the baselines grow from 0.5–0.9~s\textsuperscript{–1} to 1.8–2.7~s\textsuperscript{–1} at 10~dB, whereas our method’s false positives rise modestly from 0.12 to 0.50~s\textsuperscript{–1} at 18~dB and stabilize around 0.46~s\textsuperscript{–1} at 10~dB.

Overall, these results confirm that our method is substantially more robust to additive noise, preserving high‐fidelity reconstruction and reliable event detection under conditions that severely degrade under existing approaches.

\begin{figure}[ht]
  \centering
  \includegraphics[width=\textwidth]{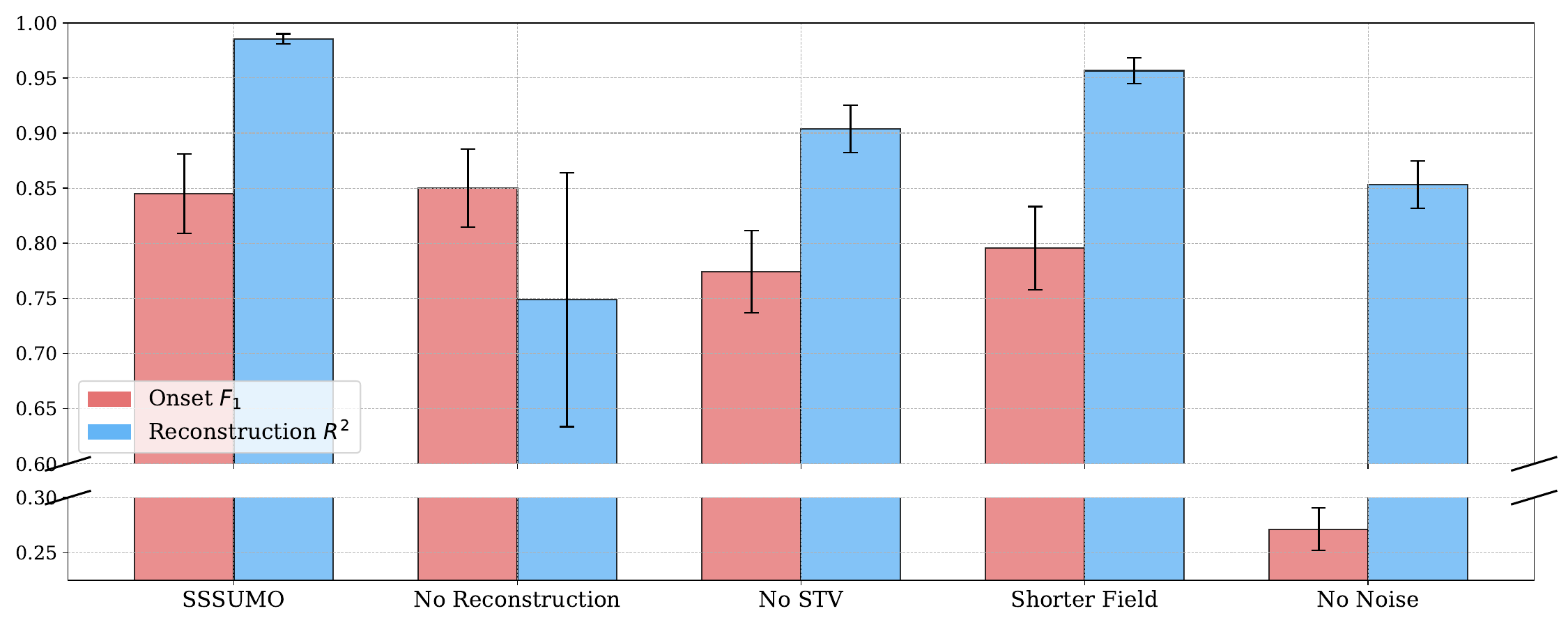}
  \caption{Ablation studies.
    Onset-detection \(F_1\) (red) and velocity-reconstruction \(R^2\) (blue) for five model variants:
    \textbf{\ssumo} (full model, receptive field $\approx$1.65 s, uses signed tangential velocity);
    \textbf{No Reconstruction} (remove reconstruction loss);
    \textbf{No STV} (use unsigned velocity instead of signed tangential velocity);
    \textbf{Shorter Field} (receptive field $\approx$0.85 s);
    \textbf{No Noise} (trained with no additive noise).
    }
  \label{fig:ablation}
\end{figure}

\paragraph{\textbf{Ablation Results.}} Figure \ref{fig:ablation} compares the full model with the modified versions.
The full model (\textbf{Ours}) achieves the highest onset‐detection \(F_1 \approx 0.85\) and velocity‐reconstruction \(R^2 \approx 0.99\). Removing the reconstruction loss (\textbf{No Reconstruction}) collapses \(R^2\) to 0.75 (interquartile range increases from 0.007 to 0.29) and yields only a marginal \(F_1\) gain of 0.005. Omitting signed tangential velocity (\textbf{No STV}) reduces onset‐detection to \(F_1 = 0.77\) and reconstruction to \(R^2 = 0.90\). Halving the receptive field (\textbf{Shorter Field}) lowers \(F_1\) to \(\approx 0.80\) and \(R^2\) to 0.96. Finally, training without additive noise (\textbf{No Noise}) drastically degrades onset detection (\(F_1 \approx 0.27\)) while maintaining moderate reconstruction (\(R^2 \approx 0.85\)). Only the full model consistently delivers both high onset accuracy and superior reconstruction fidelity.

\subsection{Human-Motion Evaluation}
\label{sec:results_human}

Synthetic benchmarks provide clear ground truth under controlled overlap and noise, but real human movements are far more variable. To assess practical performance, we evaluated our method on organic human–motion recordings spanning seven tasks and three noise conditions (no noise, 20~dB, and 10~dB; see Table~\ref{tab:organic_comparison}).

\begin{table}[ht]
    \centering
    \caption{
    Comparison of the proposed method against the benchmarks on human motion data. Average trial-wise variance explained ($R^2$) between original and reconstructed velocity signals and average predicted submovements per second (SM rate) are presented under three additive-noise conditions: no noise ($\infty$), and signal-to-noise ratios (SNR) of 20\,dB and 10\,dB. Within each task and noise level, the highest $R^2$ values are highlighted in \textbf{bold}, and the second-highest in \textit{italic}. The proposed method consistently achieves the highest $R^2$ across all tasks and noise conditions.
    }
    \begin{tabular}{l l | c c | c c | c c }
        \toprule
        \multirow{2}{*}{\textbf{Task}} & \multirow{2}{*}{\textbf{SNR (dB)}} & \multicolumn{2}{c|}{\textbf{\ssumo}} & \multicolumn{2}{c|}{\textbf{Peak Detector}} & \multicolumn{2}{c|}{\textbf{Scattershot}} \\
        \cmidrule(lr){3-4} \cmidrule(lr){5-6} \cmidrule(lr){7-8}
         &  & \textbf{$R^2$} & \textbf{SM Rate} & \textbf{$R^2$} & \textbf{SM Rate} & \textbf{$R^2$} & \textbf{SM Rate}\\
        \midrule

        \multirow{3}{*}{Steering}
         & $\infty$ & \textbf{.969}                                & 5.6                                          & \textit{.926}                                         & 2.5                                          & .886                                & 2.9                                          \\
         & 20    & \textbf{.967}                                & 4.0                                          & \textit{.923}                                         & 2.9                                          & .875                                & 3.6                                          \\
         & 10    & \textbf{.942}                           & 2.9                                          & \textit{.902}                                         & 3.7                                          & .823                                & 4.2                                          \\
        \midrule

        \multirow{3}{*}{Crank Rotation}
         & $\infty$ & \textbf{.640}                                & 4.6                                          & .011                                         & 1.8                                          & \textit{.557}                                & 2.1                                          \\
         & 20    & \textbf{.743}                                & 4.3                                          & .010                                         & 1.9                                          & \textit{.566}                                & 2.2                                          \\
         & 10    & \textbf{.797}                                & 4.8                                          & .008                                         & 2.6                                          & \textit{.599}                                & 2.3                                          \\
        \midrule

        \multirow{3}{*}{Fitts's Task}
         & $\infty$ & \textbf{.886}                                & 1.7                                          & \textit{.742}                                & 1.3                                          & .680                                         & 2.4                                          \\
         & 20    & \textbf{.887}                                & 1.5                                          & \textit{.728}                                         & 2.2                                          & .726                               & 3.4                                          \\
         & 10    & \textbf{.872}                                & 1.3                                          & .686                                         & 3.8                                          & \textit{.719}                                & 3.7                                          \\
        \midrule

        \multirow{3}{*}{Whack-A-Mole}
         & $\infty$ & \textbf{.925}                                & 5.6                                          & .676                                         & 2.4                                          & \textit{.697}                                & 2.7                                          \\
         & 20    & \textbf{.920}                                & 4.8                                          & .672                                         & 3.3                                          & \textit{.684}                                & 3.4                                          \\
         & 10    & \textbf{.893}                                & 3.0                                          & .661                                         & 4.7                                          & \textit{.692}                                & 3.5                                          \\
        \midrule

        \multirow{3}{*}{Object Moving}
         & $\infty$ & \textbf{.961}                                & 4.0                                          & .737                                         & 1.3                                          & \textit{.850}                                & 1.9                                          \\
         & 20    & \textbf{.951}                                & 2.9                                          & .742                                         & 1.6                                          & \textit{.870}                                & 2.0                                          \\
         & 10    & \textbf{.907}                                & 2.1                                          & .676                                         & 2.5                                          & \textit{.846}                                & 3.7                                          \\
        \midrule

        \multirow{3}{*}{Pointing}
         & $\infty$ & \textbf{.981}                                & 2.5                                          & .840                                         & 1.6                                          & \textit{.865}                                & 1.9                                          \\
         & 20    & \textbf{.980}                                & 1.5                                          & \textit{.833}                                & 2.7                                          & .822                                         & 2.7                                          \\
         & 10    & \textbf{.968}                                & 0.8                                          & .812                                         & 4.3                                          & \textit{.838}                                & 3.2                                          \\
        \midrule

        \multirow{3}{*}{Handwriting}
         & $\infty$ & \textbf{.864}                                & 9.1                                          & \textit{.586}                                & 4.9                                          & .503                                         & 3.6                                          \\
         & 20    & \textbf{.857}                                & 8.6                                          & \textit{.585}                                        & 5.2                                          & .499                                & 3.6                                          \\
         & 10    & \textbf{.809}                                & 7.1                                          & \textit{.568}                                & 5.6                                          & .479                                         & 3.6                                          \\
        \bottomrule   

        \multirow{3}{*}{\textbf{Mean}}
         & $\infty$ & \textbf{.890}                                & 4.7                                          & .645                                         & 2.3                                          & \textit{.720}                                & 2.5                                          \\
         & 20    & \textbf{.901}                                & 3.9                                          & .642                                         & 2.8                                          & \textit{.720}                                & 3.0                                          \\
         & 10    & \textbf{.884}                                & 3.1                                          & .616                                         & 3.9                                          & \textit{.714}                                & 3.5                                         \\
        \bottomrule         
    \end{tabular}
    \label{tab:organic_comparison}
\end{table}

Our method consistently outperforms both baselines in reconstruction accuracy across all tasks and noise levels.  The mean improvement in $R^2$ over Scattershot is 0.147, 0.162, and 0.161 for the no-noise, medium-noise, and high-noise conditions, respectively.  The margin over the Peak Detector is even larger, approximately 0.25 at each noise level.

The largest relative gain occurs on the Handwriting task. As previously reported by \citet{fast_writting}, handwriting can involve up to 12 submovements per second. In the no‐noise condition, our method detects 9.1 submovements per second (versus 4.9 and 3.6 for the Peak Detector and Optimization baseline), yielding $R^2$ values of 0.864, 0.586, and 0.503, respectively.

For all datasets except Crank Rotation, the reconstruction $R^2$ of the proposed method is only 1–4 \% lower than that produced by low‐pass filtering the original signal (Table~\ref{tab:datasets}). This shows that our approach is closing the gap toward perfect velocity reconstruction across diverse recording settings and tasks. The Crank Rotation dataset is particularly challenging for every method. This is explained by the mechanics of the task: participants rotated a crank continuously, making submovement boundaries difficult to locate. The improvement in our reconstruction with added noise suggests that, in such distant domains, fine‐grained features learned from previously seen data can mislead the model; when those finer cues are partially obscured by noise, the model relies instead on coarser structure and therefore reconstructs the signal more accurately.

We also observe that, compared with the alternative methods, the proposed algorithm detects the greatest number of submovements per second under the no‐noise condition for every dataset except the inherently noisy Fitts’s Task. Under high noise it yields the fewest detections—apart from the two most challenging datasets, Crank Rotation and Handwriting. Overall we see the same trend as in the synthetic experiments: as noise increases, our algorithm adopts a more conservative decomposition with fewer submovements, whereas the baselines show the opposite behaviour.

\begin{table}[ht]
    \centering
    \caption{Comparison of the base model with in-domain and out-of-domain fine-tuned versions on human motion tasks. Average trial-wise variance explained ($R^2$) between original and reconstructed velocity signals and average predicted submovements per second (SM Rate) are reported under three additive-noise conditions: no noise ($\infty$), 20\,dB and 10\,dB SNR. Within each task and noise level, the highest $R^2$ is highlighted in \textbf{bold} and the second-highest in \textit{italic}. In-domain fine-tuning yields substantial improvements over the base model across nearly all tasks and noise levels, while out-of-domain fine-tuning provides moderate gains, particularly under higher noise.}
    \begin{tabular}{l l | c c | c c | c c }
        \toprule
        \multirow{2}{*}{\textbf{Task}} & \multirow{2}{*}{\textbf{SNR (dB)}} & \multicolumn{2}{c|}{\textbf{Base}} & \multicolumn{2}{c|}{\textbf{In-Domain}} & \multicolumn{2}{c|}{\textbf{Out-of-Domain}} \\
        \cmidrule(lr){3-4} \cmidrule(lr){5-6} \cmidrule(lr){7-8}
         &  & \textbf{$R^2$} & \textbf{SM Rate} & \textbf{$R^2$} & \textbf{SM Rate} & \textbf{$R^2$} & \textbf{SM Rate} \\
        \midrule
        \multirow{3}{*}{Steering} 
         & $\infty$  & .964                 & 5.8                  & \textbf{.972}        & 5.5                  & \textit{.969}        & 5.6                  \\
         & 20 & .962                 & 4.3                  & \textbf{.970}        & 4.0                  & \textit{.967}        & 4.0                  \\
         & 10 & \textit{.943}        & 3.6                  & \textbf{.943}        & 2.9                  & .942                 & 2.9                  \\
        \midrule

        \multirow{3}{*}{Crank Rotation} 
         & $\infty$  & \textit{.655}        & 5.1                  & \textbf{.852}        & 5.6                  & .640                 & 4.6                  \\
         & 20 & .653                 & 4.2                  & \textbf{.915}        & 5.6                  & \textit{.743}        & 4.3                  \\
         & 10 & .555                 & 3.4                  & \textbf{.839}        & 5.7                  & \textit{.797}        & 4.8                  \\
        \midrule

        \multirow{3}{*}{Fitts's Task} 
         & $\infty$  & \textit{.888}        & 1.9                  & \textbf{.889}        & 1.9                  & .886                 & 1.7                  \\
         & 20 & \textbf{.889}        & 1.8                  & \textit{.889}        & 1.7                  & .887                 & 1.5                  \\
         & 10 & \textbf{.877}        & 1.8                  & \textit{.876}        & 1.5                  & .872                 & 1.3                  \\
        \midrule

        \multirow{3}{*}{Whack-A-Mole} 
         & $\infty$  & .911                 & 4.6                  & \textbf{.930}        & 5.6                  & \textit{.925}        & 5.6                  \\
         & 20 & .908                 & 4.2                  & \textbf{.926}        & 5.0                  & \textit{.920}        & 4.8                  \\
         & 10 & .886                 & 3.6                  & \textbf{.900}        & 3.3                  & \textit{.893}        & 3.0                  \\
        \midrule

        \multirow{3}{*}{Object Moving} 
         & $\infty$  & .957                 & 4.9                  & \textbf{.971}        & 4.0                  & \textit{.961}        & 4.0                  \\
         &  20 & \textit{.953}        & 3.1                  & \textbf{.973}        & 3.1                  & .951                 & 2.9                  \\
         &  10 & \textit{.927}        & 2.4                  & \textbf{.928}        & 2.2                  & .907                 & 2.1                  \\
        \midrule

        \multirow{3}{*}{Pointing} 
         &  $\infty$ & .976                 & 3.9                  & \textbf{.982}        & 2.7                  & \textit{.981}        & 2.5                  \\
         &  20 & .976                 & 2.0                  & \textit{.976}        & 1.5                  & \textbf{.980}        & 1.5                  \\
         &  10 & \textit{.970}        & 1.3                  & \textbf{.970}        & 0.8                  & .968                 & 0.8                  \\
        \midrule

        \multirow{3}{*}{Handwriting} 
         &  $\infty$ & .837                 & 8.5                  & \textbf{.876}        & 10.8                 & \textit{.864}        & 9.1                  \\
         &  20 & .832                 & 8.2                  & \textbf{.868}        & 10.1                 & \textit{.857}        & 8.6                  \\
         &  10 & .796                 & 7.5                  & \textbf{.815}        & 8.0                  & \textit{.809}        & 7.1                  \\
         \bottomrule

        \multirow{3}{*}{\textbf{Mean}} 
         &  $\infty$ & .884                 & 5.0                  & \textbf{.925}        & 5.2                  & \textit{.890}        & 4.7                  \\
         &  20 & .882                 & 4.0                  & \textbf{.931}        & 4.4                  & \textit{.901}        & 3.9                  \\
         &  10 & .851                 & 3.4                  & \textbf{.896}        & 3.5                  & \textit{.884}        & 3.1                  \\
        
        \bottomrule
    \end{tabular}
    \label{tab:base_model}
\end{table}

\paragraph{\textbf{Fine-Tuning Evaluation.}}
Table \ref{tab:base_model} summarises the effects of in-domain and out-of-domain fine-tuning on the proposed model. As expected, \emph{in-domain} fine-tuning boosts reconstruction performance on every dataset except Fitts's Task, where the change is negligible. Averaged across all tasks, it yields a 4–5 \% increase in $R^2$, with the largest gain observed for Crank Rotation.

\emph{Out-of-domain} fine-tuning provides more modest benefits, improving mean $R^2$ by 0.6 \%, 1.9 \%, and 3.3 \% under the no-noise, medium-noise, and high-noise conditions, respectively. The biggest gains appear on the harder tasks—most notably Handwriting (1.3–2.7 \% across noise levels) and Crank Rotation (but only when additive noise is present). As noted earlier, these improvements likely arise because fine features learned in other domains can mislead the model in unfamiliar settings; obfuscating those cues with noise forces the network to rely on coarser structure.
\begin{table}[ht]
    \centering
    \caption{Processing time and computational complexity of models for a 1-second input signal. Measured on CPU unless stated otherwise.}
    \begin{tabular}{lcc}
    \textbf{Model} & \textbf{Processing Time (s)} & \textbf{Computational Complexity} \\
    \midrule
    \textbf{\ssumo}           & $6.7\times10^{-4}$ ( $1.6\times10^{-6}$ on GPU ) & $\mathcal{O}(n)$ \\
    \textbf{Peak Detector}   & $\mathbf{3.6\times10^{-4}}$          & $\mathcal{O}(n)$ \\
    \textbf{Scattershot}   & $3.3$                       & $\mathcal{O}(n^3)$
    \end{tabular}
    \label{tab:speed}
\end{table}

\paragraph{\textbf{Processing Time Evaluation.}}
Table~\ref{tab:speed} shows that, on a CPU, our model is roughly half as fast as the Peak Detector, requiring \(6.7 \times 10^{-4}\) s to process one second of data. By contrast, Scattershot needs 3.3 s for the same input— \(\approx 5000\) times slower than our approach. When run on a GPU in batch mode, our method gets a \( \approx 400 \times \) speed-up, becoming about 200 × faster than the Peak Detector. Because Scattershot’s optimiser has polynomial-time complexity, its runtime is expected to grow cubically with signal length unless runtime-conserving heuristics (e.g., sliding-window processing) are employed.

\section{Discussion}

In this work, we demonstrate a deep learning method, \ssumo, for accurate and fast submovement decomposition. Our approach utilizes synthetic data derived from submovement principles, unlabeled organic data, a fully convolutional backbone, differentiable reconstruction, and iterative synthetic data adaptation.

We showed that submovement decomposition presents an ideal use case for semi-supervised learning with synthetic data, as the hypotheses about how submovements are generated (minimum-jerk profiles) have been extensively studied. While previous research has attempted to analytically or empirically determine optimal features for detecting submovements, our approach allows a model to `reverse-engineer' patterns directly from synthetic data generated according to established assumptions, and further refine the reconstruction performance by sequential synthetic data adaptation to real human-motion data. Rather than manually engineering features (such as wavelets or velocity peaks), we demonstrate that neural networks can effectively learn the underlying patterns that characterize submovements.

\subsection{Performance}

The \ssumo method achieved state-of-the-art performance, outperforming established baselines on both synthetic and organic benchmarks, without task-specific fine-tuning. \ssumo particularly excelled in recovering ground-truth submovement parameters (onset, duration, displacement) for synthetic data and demonstrated high reconstruction fidelity ($R^2$) on challenging Crank Rotation and Handwriting.

Our results show that out-of-domain adaptation improved the model performance particularly on the most challenging human-motion datasets, while in-domain adaptation further enhanced reconstruction performance, especially for complex tasks. Some datasets did not substantially benefit from the in-domain adaptation, e.g. Steering, suggesting performance might saturate for simpler, or less noisy tasks. The significant performance gains observed in challenging datasets suggest that our approach can effectively adapt to diverse natural movement patterns, even when those patterns diverge substantially from the synthetic training data. This adaptability is particularly important for real-world applications, where movement characteristics may vary widely across individuals, tasks, and settings.

It is worth highlighting that the proposed method does not fit directly on the human-motion data, but instead captures general statistics that are iteratively extracted to create more relevant synthetic data. Thus, the risk of overfitting is minimal, though we did perform precautionary train-test splits for evaluation. This allows future users to run adaptation on their data without concerns about double-dipping.

\subsection{Practical Applications}

Submovement analysis has become an important tool in motor control research, and more accurate real-time decomposition radically expands potential applications. Submovement decomposition is clinically important for motor disorder detection \citep{stroke_recovery2, hogan_stroke}. In healthcare settings, accurate real-time decomposition technology could be incorporated into monitoring devices for elderly populations that detect early signs of neurological conditions by identifying subtle changes in movement patterns. Early detection systems could use changes in submovement parameters to identify warning signs of stroke, or movement disorders such as Parkinson's disease or ataxia, before they become clinically apparent. In rehabilitation, it could provide real-time feedback on movement quality and progress. 

Beyond clinical applications, systems could dynamically assess task difficulty for users, applicable to adaptive interfaces \citep{autogain, interface1, interface2, robot} or safety-critical applications, e.g. driving \citep{markkula_control}. \ssumo could enhance human-computer interaction through more accurate intention detection, enabling interfaces that adapt to users' movement capabilities. For sports training, coaches could use detailed submovement analysis to identify technical inefficiencies that are invisible to the naked eye.



\subsection{Limitations and Future Improvements}

\paragraph{\textbf{Inference-time optimization.}}
While \ssumo provides highly accurate decompositions, it still makes mistakes, particularly when relying on local features that can be misleading in novel settings. Such errors could potentially be addressed by initializing local optimization with model outputs. A more holistic approach would be to allow the model to iteratively incorporate its own outputs and reconstruction results, enabling it to use previous submovement estimates to improve subsequent predictions. This would enhance understanding of the interdependency between submovement parameters and the resulting velocity signal. This approach relates to test-time computation \citep{test_time} and loop networks \citep{loop}, allowing smaller models to benefit in accuracy from increased runtime. 


This iterative refinement approach offers an additional advantage in terms of context utilization. Since each output element is calculated based on the model's receptive field, feeding this output back into the model effectively expands the contextual information available for subsequent predictions. This implicit encoding of broader temporal context could lead to more coherent and accurate submovement decompositions across longer movement sequences.

\paragraph{\textbf{Incorporating global features.}}
Our fully convolutional architecture, relying only on local features, significantly outperforms existing methods. However, our ablation studies show that decreasing the receptive field substantially reduces performance. Thus, increasing the receptive field, perhaps by combining the convolutional architecture with recurrent or attention blocks, could potentially further improve \ssumo's capabilities, albeit with increased computational overhead.

\paragraph{\textbf{Learnable submovement velocity profiles.}}
Another potential extension relates to learnable shapes of submovement velocity profiles. The commonly used Min-Jerk-Velocity profile is parameterized by displacement and duration. However, ongoing research investigates whether shapes described by additional parameters should be considered, as human motor systems may execute different velocity profiles \citep{spurious2, beta}.

For example, normalized min-jerk-velocity can be generalized using the Beta density function, using $\alpha$ and $\beta$ parameters to characterize peakness and skewness \citep{beta}. Simply optimizing such parameters for signal decomposition can lead to ambiguous decompositions, as it increases the space of possible solutions. It becomes unclear whether a velocity profile should be described as a combination of symmetric submovements or one larger, skewed submovement.

In contrast, our architecture allows for a novel approach: while the detector module predicts onset, duration, and displacement parameters, the reconstruction module could be augmented with learnable submovement shapes. These could take the decomposed parameters alongside additional features (e.g., task type, subject ID, experience and difficulty levels) to regress submovement parameters leading to the most accurate reconstruction. This method would allow for generalizing submovement parameters based on learned dependencies within movement descriptions, potentially revealing factors that affect submovement shapes.

\paragraph{\textbf{Improved real-world data adaptation.}}
Kernel density estimation (KDE) estimation allows us to estimate the distribution of submovements in organic data which leads to performance gains, but it is computationally costly when modeling multi-dimensional conditional probabilities. Normalizing flows \citep{norm_flows} could provide a more efficient alternative for approximating these distributions, offering better scalability while maintaining modeling power.

Additionally, organic data patterns could be imitated using generative adversarial networks (GANs) \citep{gan} or other generative models. This approach would allow us to synthesize realistic movement data with known ground truth, further enriching the training process and potentially improving generalization to novel movement patterns.

An additional consideration is how noise is generated in the signals we utilize. Currently, we apply Gaussian additive noise, but in practice, we observe distinct noise patterns in human motion data that mainly arise from quantization and jitter errors of the sensor \citep{noise}. Modeling these specific noise characteristics could further improve \ssumo's robustness in real-world settings.

\paragraph{\textbf{Dimensional limitations}}
A significant limitation of \ssumo is that it does not work natively with higher-dimensional movement data. While converting spatial velocity signals to STV preserves binary information about whether subsequent submovements reinforce or counteract each other, it loses positional information and doesn't allow unambiguous reconstruction of spatial velocity signals. Passing multi-dimensional velocity as input could reduce submovement ambiguity, as submovements would possess distinct directions rather than merely overlapping in a 1D signal.

Among the tested datasets, Crank Rotation proved the most challenging for all methods. While \ssumo with in-domain fine-tuning achieved a high reconstruction $R^2$ of 0.92 on 20 dB SNR, this remains significantly below optimal performance as shown in Table \ref{tab:datasets}. This raises questions about whether such motion is inherently impossible to decompose, highlighting potential limits of submovement decomposition, or whether alternative approaches are needed -- perhaps reconstructing spatial representations of crank rotation, where spatial arrangements of submovements might facilitate their disambiguation.

\paragraph{\textbf{Sampling rate considerations.}}
While 60 Hz appears to be a reasonable sampling rate for preserving features necessary for submovement detection, the model could benefit from handling signals of various sampling rates to consider more granular features in higher-sampled signals.

\paragraph{\textbf{Expanding the scope of the method.}}
\ssumo can be beneficially combined with modern pose estimation models for video data. This integration potentially allows extraction of submovement parameters from video data without the need for additional sensors, dramatically expanding accessibility for research and applications. An exciting direction for future work involves extending the method to tasks with multiple effectors -- such as two hands in typing or playing piano, arms and legs in complex sports movements, steering wheel and pedals in driving, or eye-head and eye-hand coordination tasks. These multi-effector scenarios present unique challenges in terms of temporal coordination and spatial dependencies between movement components.

\subsection{Broader Implications for Motor Control Research}

The success of \ssumo in noisy natural datasets supports the idea, mentioned in section~\ref{sec:intro}, that submovement detection can be used to infer sensorimotor system strategies. Neurophysiological studies suggest that discrete submovements may be linked to intrinsic oscillatory dynamics in the motor cortex, typically in the 2–4 Hz range \citep{neural_evidence1, neural_evidence2, neural_SMA, fast_neuro}, providing additional support for the intermittent nature of motor control. 

There is ongoing debate regarding whether movements are intrinsically composed of discrete submovements \citep{markkula_control, control2, control4, review_paper1}, or if such segmentation merely reflects an emergent property of continuous control systems \citep{opposite_view1, opposite_view2, opposite_view3}. The present work does not aim to test the hypothesis of intermittent control managed via submovements, but solely to provide a method for more accurate and fast decomposition, with high application utility \citep{stroke_recovery2, game1, autogain}. Yet, the resulting model is not solely of interest for decomposition. 

To accurately detect submovement onset and characteristics, the model must capture signal dynamics in its hidden layer outputs. These outputs store information about trends and significant changes in velocity signals, allowing use of the model (with output layers removed) for continuous motor control research. This ability to extract meaningful features from movement data could facilitate new approaches to studying motor control processes. For instance, the learned representations could be analyzed to identify latent factors that explain variance across different movement types or individuals. These representations might also reveal how the motor system adapts to changing task demands or constraints, providing insights into the hierarchical organization of motor planning and execution.

Furthermore, the proposed method's computational efficiency makes it suitable for integration into closed-loop experimental paradigms, where real-time analysis of movements could inform adaptive perturbations or feedback. Such approaches could help resolve the debate about whether movements are planned as discrete units versus continuous control processes.

We make our code and the model weights available at \href{https://github.com/dolphin-in-a-coma/sssumo}{\texttt{github.com/dolphin-in-a-coma/sssumo}}.

\section*{Declarations}

\textbf{Funding}
Funding was provided by the Research Council of Finland (Grant Nos. 334192, 355200, and 361200).

\appendix

\section{Submovement Discrimination Limit}


\begin{figure}[ht!]
    \centering
    \label{fig:noise_decomposition}
    
    \begin{subfigure}{0.44\textwidth}
        \includegraphics[width=\linewidth]{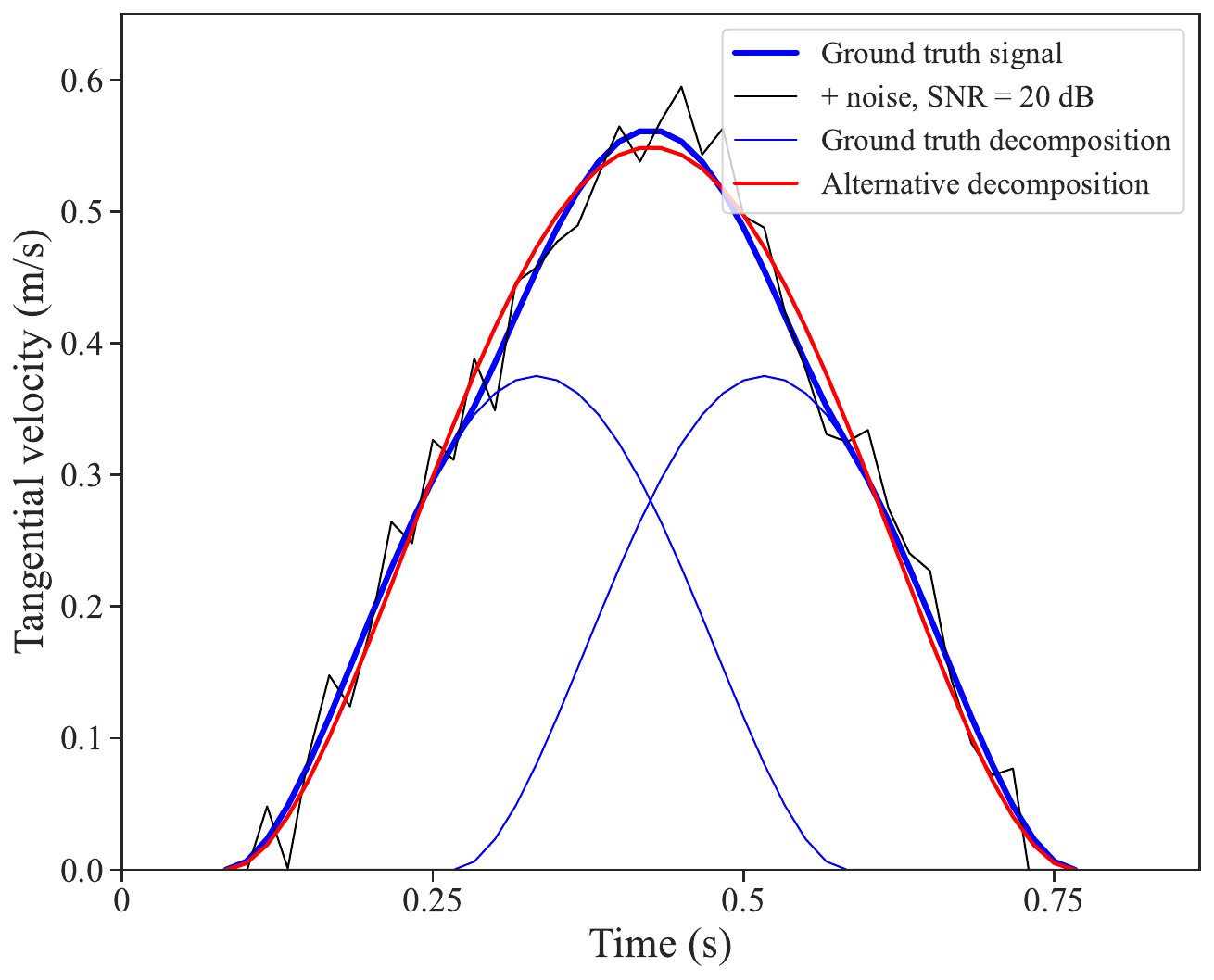}
        \caption{Decomposition of noisy signal}
    
    \end{subfigure}
    \begin{subfigure}{0.44\textwidth}
        \includegraphics[width=\linewidth]{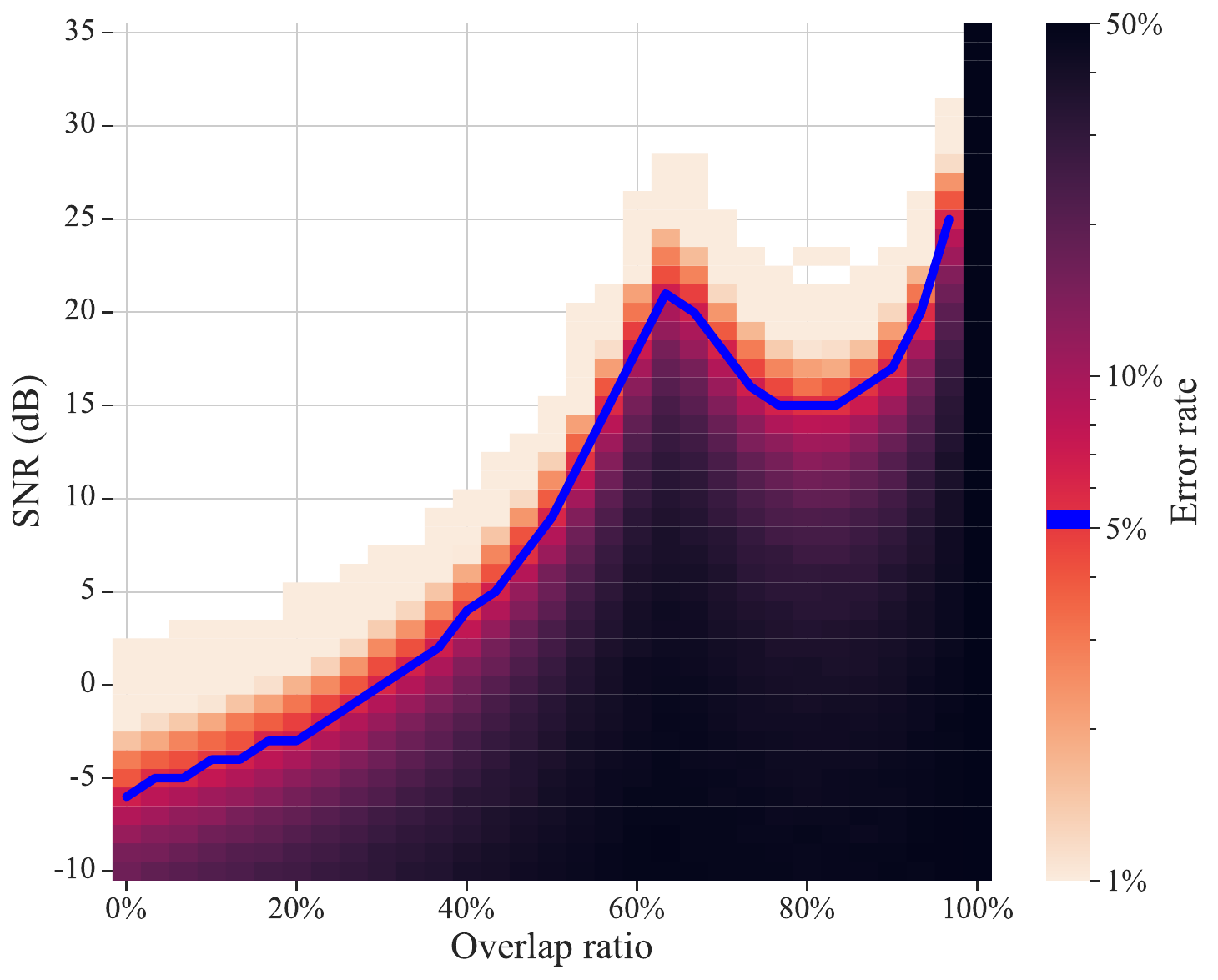}
        \caption{Discriminability of overlapping submovements}
    
    \end{subfigure}
\caption{Impact of noise on the accuracy of submovement decomposition. 
\textbf{(a)} The plot illustrates a clean velocity signal (bold blue line) alongside a noisy version (black line), generated from a superposition of two minimum jerk velocity profiles (thin blue lines) with a displacement of 0.1 m and a duration of 0.5 s, overlapping for 0.32 s (63\% of one submovement’s duration). The red line represents an alternative decomposition, modeling the signal with a single submovement of 0.68 s and a displacement of 0.2 m. At a signal-to-noise ratio (SNR) of 20 dB, the discriminability error rate between these decompositions reaches 5\%.
\textbf{(b)} Heatmap depicting the discriminability error rate as a function of the overlap ratio (relative to a single submovement's duration) and SNR. The duration and displacement as in (a). For each combination of SNR and overlap ratio, 10,000 noise simulations were performed, and the error rate was computed as the proportion of simulations where the alternative single-submovement decomposition yielded a lower mean absolute error (MAE) than the ground-truth decomposition with two overlapping submovements. The blue contour marks the 5\% error rate threshold. An overlap above 57\% and an SNR below 15 dB consistently yield a 5\% error rate.
}

\end{figure}

\section{Peak Detector Algorithm}

\begin{algorithm}[H]
\footnotesize
\caption{Peak Detector Algorithm Pseudocode}
\label{fig:peak_detector}
\begin{algorithmic}[1]
\Require Input signal $v(t)$ for $t = 1, 2, \dots, N$; Threshold $\theta$ (default: 0.01) ; Smoothing parameter $\sigma$ (default: 2)
\Ensure List of submovements, each with onset, peak, offset, duration, and amplitude (AUC)

\State Initialize empty lists: \texttt{peaks}, \texttt{onset\_candidates}, \texttt{submovements}

\Comment{\textbf{Step 0: Smooth the Input Signal}}
\State $v_{smooth}(t) \gets \text{GaussianSmooth}(v(t), \sigma)$ \Comment{Apply Gaussian smoothing with parameter $\sigma$}

\Comment{\textbf{Step 1: Identify Absolute Peaks (Submovement Peaks) and Valley Extremes (Onset Candidates )}}
\For{$t = 2$ to $N-1$}
    \If{$v_{smooth}(t-1) < v_{smooth}(t) \land v_{smooth}(t) > v_{smooth}(t+1)$} \Comment{Local maximum}
        \If{$v_{smooth}(t) > \theta$} \Comment{Positive maximum}
            \State \texttt{peaks.append($t$)}
        \ElsIf{$v_{smooth}(t) < 0 $} \Comment{Negative maximum}
            \State \texttt{onset\_candidates.append($t$)}
        \EndIf
    \ElsIf{$v_{smooth}(t-1) > v_{smooth}(t) \land v_{smooth}(t) < v_{smooth}(t+1)$} \Comment{Local minimum}
        \If{$v_{smooth}(t) < \theta$} \Comment{Negative minimum}
            \State \texttt{peaks.append($t$)}
        \ElsIf{$v_{smooth}(t) > 0 $} \Comment{Positive minimum}
            \State \texttt{onset\_candidates.append($t$)}
            
        \EndIf
    \EndIf
\EndFor

\Comment{\textbf{Step 2: Identify Zero-Crossing Events and Add to Onset Candidates}}
\For{$t = 1$ to $N-1$}
    \If{$v_{smooth}(t) = 0$ \textbf{or} $v_{smooth}(t) \times v_{smooth}(t+1) < 0$}
        \State \texttt{onset\_candidates.append($t$)}
    \EndIf
\EndFor

\Comment{\textbf{Step 3: Identify Signal Stability Events and Add to Onset Candidates}}
\State $is\_stable \gets False$
\For{$t = 1$ to $N$}
    \If{$|v_{smooth}(t)| \leq \theta$}
        \If{$is\_stable = False$} \Comment{First stable time point}
            \State \texttt{onset\_candidates.append}($t$) \Comment{Record first stable time point}
            \State $is\_stable \gets True$
        \EndIf
    \Else
        \If{$is\_stable = True$} \Comment{Last stable time point}
            \State \texttt{onset\_candidates.append}($t-1$) \Comment{Record last stable time point}
            \State $is\_stable \gets False$
        \EndIf
    \EndIf
\EndFor



\Comment{\textbf{Step 4: Offset Candidates are the Same as Onset Candidates}}

\Comment{\textbf{Step 5: For Each Submovement Peak, Find Closest Onset and Offset Candidates, and Calculate Duration and Amplitude}}
\For{\textbf{each} $peak$ \textbf{in} \texttt{peaks}}
    \State $onset \gets \max\{ t \in \texttt{onset\_candidates} \mid t \leq peak \}$
    \State $offset \gets \min\{ t \in \texttt{onset\_candidates} \mid t \geq peak \}$
    
    \State $duration \gets offset - onset$
    
    \State $amplitude \gets \sum_{t=onset}^{offset} v(t)$ \Comment{Sum of values within the submovement interval}

    \State \texttt{submovements.append}($\{onset, peak, offset, duration, amplitude\}$)
\EndFor

\State \Return \texttt{submovements}
\end{algorithmic}
\label{algo:heur}
\end{algorithm}




\bibliography{references}

\end{document}